\documentclass[%
pre, amsmath,amssymb,showkeys,
aps,superscriptaddress,onecolumn,a4paper,english]{revtex4}
\usepackage[T1]{fontenc}
\usepackage[utf8]{inputenc}
\usepackage{amsmath}
\usepackage{amssymb}
\usepackage{graphicx}
\usepackage{bm}
\usepackage{multirow,array}
\usepackage{mathtools}
\usepackage{empheq}
\usepackage{chessfss}
\usepackage{epsdice}
\usepackage{pifont}
\usepackage{amsthm}
\usepackage[
citecolor=blue,
colorlinks,
linkcolor=blue,
urlcolor=blue,
]{hyperref}
\usepackage[dvipsnames]{xcolor}
\usepackage{dcolumn}
\usepackage{bm}
\usepackage{etoolbox}  
\usepackage{soul} 
\usepackage{cancel}
\begin{document}

	\title{Oscillatory equilibrium in asymmetric evolutionary games: Generalizing evolutionarily stable strategy}
	\author{Vikash Kumar Dubey}
	\email{vdubey@iitk.ac.in}
	\affiliation{
		Department of Physics,
		Indian Institute of Technology Kanpur,
		Uttar Pradesh 208016, India
	}
	\author{Suman Chakraborty}
	\email{suman@iitk.ac.in}
	\affiliation{
		Department of Physics,
		Indian Institute of Technology Kanpur,
		Uttar Pradesh 208016, India
	}	
	\author{Sagar Chakraborty}
	\email{sagarc@iitk.ac.in}
	\affiliation{
		Department of Physics,
		Indian Institute of Technology Kanpur,
		Uttar Pradesh 208016, India
	}

\date{\today}
	\begin{abstract}
	The concept of evolutionarily stability and its relation with the fixed points of the replicator equation are important aspects of evolutionary game dynamics. In the light of the fact that oscillating state of a population and individuals (or players) of different roles are quite natural occurrences, we ask the question how the concept of evolutionarily stability can be generalized so as to associate game-theoretic meaning to oscillatory behaviours of players asymmetrically interacting, i.e., if there are both intraspecific and interspecific interactions between two subpopulations in the population. We guide our scheme of generalization such that the evolutionary stability is related to the dynamic stability of the corresponding periodic orbits of a time-discrete replicator dynamics. We name the generalization of evolutionarily stable state as two-species heterogeneity stable orbit. Furthermore, we invoke the principle of decrease of relative entropy in order to associate the generalization of evolutionary stability with an information-theoretic meaning. This particular generalization is aptly termed as two-species information stable orbit.
	\end{abstract}
	\maketitle

{\color{black}\section{Introduction}
Dawkins~\cite{dawkins1989} remarks
that the concept of the evolutionarily stable strategy (ESS)~\cite{Maynard_Smith_Price_1973} is one of the most important advances in evolutionary theory since the theory of natural selection~\cite{darwin1859,wallace1869}. ESS, which is at the core of the evolutionary game theory~\cite{Maynard_Smith_1982}, is essentially a strategy adopted by an entire monomorphic population such that when invaded by a tiny fraction of mutants, ESS either renders a higher expected payoff through its performance against itself; or in case of a tie, it fetches more expected payoff against the mutant compared to what the mutant would. The expected payoff may be interpreted as fitness in this context. Due to the remarkable fact that ESS serves as refinement of Nash equilibrium (NE)~\cite{Nash1950, Nash1951}, the concept of ESS has found applications beyond the realm of biology, e.g., in economics~\cite{alchian1950, friedman1953, Nelson2002, Samuelson2002} and other social sciences~\cite{Sethi1996,Ohtsuki2004}.

Technically speaking, ESS is an equilibrium concept whose existence can be deduced from the knowledge of fitness values. However, how ESS is reached in a population requires knowledge about the dynamics of the frequencies of traits or phenotypes. These considerations lead to some beautiful connections between the fields of nonlinear dynamics and game theory. Specifically, the asymptotic outcomes in the dynamics of microevolution---paradigmatically modeled by the replicator equation~\cite{taylorjonker1978}---and game-theoretic concepts, viz., NE and ESS, correspond to each other: Folk theorem~\cite{Cressman_Tao_2014} propounds some such correspondences. Of particular interest to us is the fact that ESS corresponds to locally asymptotically stable fixed points of replicator equations~\cite{taylorjonker1978,hofbauer_book}.

The aforementioned connection attains a third dimension when one realizes that the game-theoretic interpretations of the fixed points can further be supplemented with interesting information-theoretic connection: Attainment of ESS in the course of time-evolution of frequencies is manifested~\cite{baez_entropy} as decrease in an appropriately constructed Kullback--Leibler (KL) divergence (also called relative entropy)~\cite{kullback1951,cover1999}. In the context of this paper, let us bring a rather obvious point to the fore: Both the game-theoretic and the information-theoretic meanings are associated with the fixed points of the replicator dynamics as ESS, by construction, corresponds to a single population state.

\subsection{Motivation}
However, in nature, we don't witness merely convergent outcomes: Oscillations are ubiquitous. To mention a few, Hori~\cite{1993_Hori} observed oscillations in the frequencies of two cichlid fish's phenotypes. Similarly, Sinervo and Lively~\cite{1996_Sinervo,2000_Sinervo} identified a rock-paper-scissors dynamic in a population of side-blotched lizards where the dynamics leads to oscillations in the frequencies of different throat-color phenotypes. In addition to the experimental findings, several recent theoretical works have underscored the significance of this topic in recent years~\cite{Weitz2016,Lin2019,Tilman2020,Yan2021,SohelMondal2024} in the context of eco-evolutionary games where influence of environment on game (and vice versa) is considered. Interestingly, not only periodic oscillations but chaotic oscillations are also possible; e.g., Sato et al.~\cite{Sato2002} showed that in the setting of two-player rock-paper-scissors game, Hamiltonian chaos can arise.

Therefore, it becomes a natural question to ask, how the three-way connection between dynamical system theory, game theory, and information theory appears in the case of oscillations. This was the focus of some recent works~\cite{archan_periodic,Bhatacharjee_etal_2023}, where the ESS concept in \emph{symmetric games} was extended to account for oscillatory outcomes. One paper~\cite{archan_periodic} tapped into the ideas of dynamical stability to introduce the concept of heterogeneity factor, which is multiplied with the payoff, and ESS is extended to become Heterogeneity Stable Orbit (HSO) which corresponds to oscillations optimizing the heterogeneity-weighted payoff. Interestingly, the HSO emerged as a necessary condition for stable periodic orbits. On the other hand, the other paper~\cite{Bhatacharjee_etal_2023} takes a different approach---utilizing information theory, specifically KL-divergence, as a guiding principle to extend the ESS concept to periodic orbits. This extension is known as Information Stable Orbit (ISO). Additionally, some studies~\cite{Rand_1994,archan_chaos} attempted to understand chaos from an evolutionary game theory perspective. However, none of the above mentioned papers explore \emph{asymmetric games} and addressing this lacuna is the primary focus of our paper.

A multiplayer game is called asymmetric game when the players have certain roles and with each role a set of strategies is associated. The interaction between players of same role is called intraspecific and the interaction between players of different roles is called interspecific. In this paper, we focus on a population having only two-role-two-player games with expected payoffs extractable from payoff matrices. The truly asymmetric games, i.e., in the complete absence of  intraspecific interactions, are also aptly called \emph{bimatrix games}. There have been multiple attempts to define ESS in asymmetric games. Notable are the attempts by Pohley et al.~\cite{Pohley1979}, Selten~\cite{Selten1980}, Schuster et al.~\cite{Schuster_1981}, and Hofbauer et al.~\cite{hofbauer_book} in the case of \emph{bimatrix games} (where intraspecific interaction is absent); and by Taylor~\cite{Taylor_1979} and  Cressman et al.~\cite{cressman_1992,Cressman1996,Cressman_Tao_2014} in the general case of \emph{asymmetric games} (involving both inter- and intra-specific interactions). 

A subtle and important point to highlight is that in the context of bimatrix games, mixed ESS is not possible if the definition of ESS as given by Selten~\cite{Selten1980,Schuster_1981} is adopted. Hence, extending this definition to periodic orbits in \emph{bimatrix games} would  be inappropriate, since periodic orbits in replicator dynamics must inherently consist of mixed states---after all, any pure state corresponds directly to fixed points of the replicator dynamics. While many similarities and dissimilarities between various ESS definitions~\cite{Pohley1979,Selten1980,Schuster_1981,hofbauer_book,cressman_1992,Cressman1996,Cressman_Tao_2014} (also see a contextual review~\cite{Dubey_etal2024_arXiv} on this issue) exist, for the purpose of this paper, it suffices to focus on the two-species ESS (2ESS)---an extension of ESS---proposed by Cressman et al.~\cite{cressman_1992,Cressman1996,Cressman_Tao_2014}. As far as this paper is concerned, some very important properties of 2ESS are that mixed 2ESS is possible and it is a locally asymptotically stable interior fixed point of a time-continuous replicator equation.

Having succinctly delineated the necessary background literature on the topic, we now outline the objectives of this paper. Given its technical nature, this also serve as a summary of the main new results.

\subsection{Objective and Summary}
The examples of oscillatory outcomes we presented earlier are not just confined to a single species. They have been observed even when the interaction is between multiple species, whether it is the animal kingdom~\cite{Bellani2020,Yoshida2003} or a host-parasite systems~\cite{Dybdahl1998}. This real life phenomenon immediately raises the question: How do we extend the concept of ESS to accommodate oscillatory outcomes in this context? What tools and techniques should we use to establish the three-fold connection between dynamical system theory, game theory and information theory? We achieve our goal through two approaches: the first one is based on dynamical stability~\cite{Pandit2018,Cressman_Tao_2014,Harper2014,Yoshioka2024} resulting in what we term the two-species Heterogeneity Stable Orbit (2HSO); and the second one is based on the minimization of relative entropy~\cite{baez_entropy,bomze1991cross,karev2010,Olivares2007,Qian1991,Floerchinger2020}, which we refer to as the two-species Information Stable Orbit (2ISO).

In the former approach, we demonstrate that locally asymptotically stable periodic orbits can indeed represent evolutionarily stable outcomes. However, this requires an appropriate generalization and redefinition of the concepts of 2ESS and NE to account for periodic outcomes. By doing so, we ensure that individuals in the population optimize their strategies to achieve an evolutionarily optimal `fitness'. In this generalized game setting however, the effective `fitness' must be (re-)interpreted as the product of an individual's fitness and the probability that two randomly chosen members of the population belong to different phenotypes~\cite{archan_periodic}. Populations achieving this optimizated redefined fitness state are said to be in 2HSO (two-species Heterogeneity Stable Orbit).

For the second approach we notice that, Darwinian evolution optimizes the amount of biological information generated during the process, guiding the system toward an equilibrium state~\cite{karev2010}. This principle is observed as a universal law across various fields~\cite{Olivares2007,Qian1991,Floerchinger2020}. In the context of evolution, the minimization of relative information (KL-Divergence) serves as a mechanism that drives the system toward equilibrium~\cite{baez_entropy,bomze1991cross,karev2010}. Building on this guiding principle, we extend the concept of 2ESS to periodic orbits in asymmetric games, introducing the 2ISO (two-species Information stable orbit) framework. To complete the three-fold connection we furthermore establish a relation between 2HSO and 2ISO.

Now the important issue is what dynamical model we must use given the existence of plethora of dynamical models available in the literature~\cite{taylorjonker1978,Vilone2011,Pandit2018,Gilboa1991,Brown1951,Smith1984,Blume1993,Nagurney1997} to arrive at 2HSO and 2ISO. To this end, we use the asymmetric version of the time-discrete type-I replicator map~\cite{Vilone2011}. The primary reason is that it has some desirable properties, viz., (i) it is compatible with the Darwinian tenet of natural selection, (ii) its fixed points correspond to NE and 2ESS, and (iii) most desirably, it is endowed with attracting periodic orbits. Moreover, one may envisage~\cite{Dubey_etal2024_arXiv} a correspondence between dynamical stability and 2ESS in asymmetric games just like it happens in symmetric games~\cite{Pandit2018} via replicator map. A pragmatic reason of working with the time-discrete version of replicator dynamics is that the map renders analytical tractability since, unlike its continuous-time counterpart, its periodic orbit is not a closed continuous curve but a set of countable phase points that appear even in one dimensional phase space~\cite{Hofbauer1981,hofbauer_book, Vilone2011,Pandit2018}. 

However, it would not be correct to conclude that the type-I replicator map is just a toy model---an Euler discretization of the continuous time equation; it has an independent existence and significance as highlighted by the following points: (i) It can be derived from a recursion for viability selection in two-type systems~\cite{mcelreath2008mathematical} which provides a microscopic foundation for the replicator map; (ii) while differential equations are appropriate for overlapping generations, discrete equations are more suitable for non-overlapping generations~\cite{weibull1997}; (iii) it has been used to model intergenerational cultural transmission~\cite{Montgomery2010}, the imitative behavior of boundedly rational players in bimatrix cyclic games~\cite{Hofbauer2000} and reinforcement learning~\cite{Brgers1997,Bisin2000}, which shows its relevance extends beyond traditional evolutionary contexts in biological systems; and (iv) along with periodicity, it also shows chaos~\cite{archan_chaos,Bhatacharjee_etal_2023}---another interesting avenue to look into.

Let us now outline how the rest of the paper is arranged. To clearly implement our goal, in Section~\ref{sec:convergent_outcomes}, we begin by mathematically defining our system and revisiting the existing definitions of two-species ESS, Nash equilibrium, and replicator dynamics for asymmetric games. In Section~\ref{sec:III}, before introducing the concepts of Heterogeneity Orbit (HO) and 2HSO which extend Nash equilibrium and two-species ESS, respectively, for periodic orbits, we address an important technical aspect of replicator map (Section~\ref{sec:tdre}): The dynamics of replicator map need not always stay within the physically allowable region for all parameter values making it crucial to identify the parameter values that ensure that the variables remains inside the physical region at all times. We then prove a key proposition that links 2HSO to asymptotically stable periodic orbits of the replicator map. Finally, we conclude the section by relating 2HSO to concept of strong stability~\cite{hofbauer_book}. Our information-theoretic exploration begins with proving that the decrease in KL-Divergence leads to 2ESS in Section~\ref{sec:info_fixed}, which we later extend to 2ISO in Section~\ref{sec:info_periodic}. We end this part by connecting 2ISO with asymptotic stability in Section~\ref{iso_dynamics}. Before concluding our results in Section~\ref{sec:VI}, we provide numerical exposition and validation of various newly introduced concepts through carefully chosen examples in Section~\ref{sec:numver}.
	
}

\section{Game-theoretic Equilibria and Convergent Dynamic Outcomes}\label{sec:convergent_outcomes}

In this section, we begin by examining the convergent outcomes, which indicates the stable fixed points of the corresponding dynamics, in asymmetric games. But before delving into the mathematical intricacies, we provide a preliminary exposition of the underlying framework for asymmetric games tuned to our purpose.

Suppose we have a population game involving individuals who assume one of two distinct roles, such as father/mother for parental conflicts,  owner/intruder for invasion conflicts, or host/parasites. In these games, there can be two types of interactions: one between individuals of the same role, called intra-specific interaction, and another between individuals of different roles, called inter-specific interaction. Let us confine ourselves to the cases where the payoff of any strategy is a linear function of the (components of) state of the population. We define \emph{bimatrix games} as the one with only the inter-specific interaction present. On the other hand, we term a generalization of bimatrix game as \emph{asymmetric game} that is defined to have both inter- and intra-specific interactions simultaneously. 

We analyze an underlying normal form game comprised of ${N}_{11}$ pure strategies for the role 1 player when playing against role 1, and ${N}_{12}$ pure strategies when playing with role 2 player. Similarly, for the role 2 player, the underlying normal form game has ${N}_{21}$ pure strategies against role 1 and ${N}_{22}$ pure strategies against itself. Consequently, we have two payoff matrices for the role 1 player: ${\sf U}_{11}$ of size $N_{11} \times N_{11}$ and ${\sf U}_{12}$ of size $N_{12} \times N_{21}$. Similarly, we can define the payoff matrices for role 2 individuals as ${\sf U}_{21}$ of size $N_{21}\times N_{12}$ and ${\sf U}_{22}$ of size $N_{22}\times N_{22}$. Any mixed strategy for role 1 belongs to the $\Sigma_{N_{11}}\times \Sigma_{N_{12}}$ space, and any mixed strategy for role 2 belongs to the $\Sigma_{N_{21}}\times \Sigma_{N_{22}}$ space. Here, the symbol $\Sigma_N$ denotes the simplex of dimension $N-1$.

Next, let us consider two interacting infinite-sized, well-mixed, and unstructured populations: There are $\alpha$-(pheno)types of population 1 with frequencies $x_1,x_2,\cdots,x_{\alpha}$ such that every type can be mapped to a particular strategy $\mathbf{p}_i\in\Sigma_{N_{11}}$ and another one $\mathbf{p}_{i}'\in\Sigma_{N_{12}}$. Similarly, we have $\beta$-(pheno)types in population 2 with frequencies $y_1,y_2,\cdots,y_{\beta}$ such that every type can be mapped to particular mixed strategies, $(\mathbf{q}'_j,\mathbf{q}_j)$ where $\mathbf{q}'_j\in\Sigma_{N_{21}}$ and $\mathbf{q}_j\in\Sigma_{N_{22}}$. It goes without saying that frequencies are non-negative real numbers between $0$ to $1$ and they are normalized such that they add up to $1$ at all times. {\color{black} Note that, for clarity and generality of the definitions, we are allowing arbitrary values of $\alpha$ and $\beta$ here. However, in some of the later proofs, we simplify the case to two types ($\alpha=\beta=2$) only. This simplification is motivated by two reasons: first, the qualitative insights remain the same; and second, extending the proofs to accommodate larger number of types sometimes is significantly more clumsy, making it less suitable for clear presentation of our ideas.}

Let an element ${\sf A}_{ij}$ of  matrix ${\sf A}$ be given by $\mathbf{p}_i\cdot {\sf U}_{11}\mathbf{p}_j$ and an element ${\sf B}_{ij}$ of  matrix ${\sf B}$ be given by $\mathbf{p}'_i\cdot {\sf U}_{12}\mathbf{q}'_j$. Obviously, ${\sf A}$ is the payoff matrix of an individual of population 1 and it represents the payoffs corresponding to the binary interaction between two types of population 1; ${\sf B}$ is the payoff matrix of an individual of population 1 while engaged in the binary interaction with another individual of population 2.  Similar interpretation is associated with the payoff matrices ${\sf C}$ and ${\sf D}$ with elements  ${\sf C}_{ij}=\mathbf{q}'_i\cdot {\sf U}_{21}\mathbf{p}'_j$ and  ${\sf D}_{ij}=\mathbf{q}_i\cdot {\sf U}_{22}\mathbf{q}_j$, respectively, that correspond to the individuals of population 2. It helps to define $\mathbf{x}\equiv(x_1,x_2,\cdots,x_{\alpha})^T$ and $\mathbf{y}\equiv(y_1,y_2,\cdots,y_{\beta})^T$---also called \emph{state vectors} of population 1 and 2, respectively. The overall state of the system is, thus, the state pair $(\mathbf{x},\mathbf{y})\in \Sigma_{\alpha}\times \Sigma_{\beta}$. Furthermore, whenever explicitly required, we denote the states of population 1 and population 2 at time $t$ as $\mathbf{x}^{t}$ and $\mathbf{y}^{t}$,  respectively.

Given any set of matrices ${\sf A,\,B,\,C,\,D}$, there is a well-defined prescription of what strategies von Neumann--Morgenstern rational players~\cite{von_Neumann_Morgenstern_1944} must normatively play such that no player deviates from her played strategy. Such an equilibrium known as Nash equilibrium (NE)~\cite{Nash1950,Nash1951}, mathematically reappears in the setting of Darwinian evolution~\cite{darwin1859} where rationality is replaced by natural selection: Certain NEs, calculated from the mere knowledge of the matrices turn out to be Evolutionarily Stable Strategies (ESS)~\cite{Maynard_Smith_Price_1973,Maynard_Smith_1982}. An ESS, when adopted by the host population, is, in principle, immune to invasion by an alternative infinitesimal mutant strategy. While we will present these concepts in their rigorous mathematical form as we go along, what we are trying to allude to here is that game-theoretic equilibria, e.g., NE and ESS, can be calculated without any knowledge of the dynamics of the state of the populations. 

For asymmetric games, NE is defined as a state profile that is the best response to each other. In mathematical notation, the state pair $(\mathbf{\hat{x},\hat{y}}) \in \Sigma_\alpha \times \Sigma_\beta$ is NE if the following simultaneously hold~\cite{Schuster_1981_III}:
	\begin{eqnarray}\label{11}
		\mathbf{\hat{x}}\cdot{\sf A} \mathbf{\hat{x}}+ \mathbf{\hat{x}}\cdot{\sf B} \mathbf{\hat{y}}\ge \mathbf{x}\cdot{\sf A} \mathbf{\hat{x}}+ \mathbf{x}\cdot{\sf B} \mathbf{\hat{y}}~~\text{and}~~ 
		\mathbf{\hat{y}}\cdot{\sf C} \mathbf{\hat{x}}+ \mathbf{\hat{y}}\cdot{\sf D} \mathbf{\hat{y}}\ge\mathbf{y}\cdot{\sf C} \mathbf{\hat{x}}+ \mathbf{y}\cdot{\sf D} \mathbf{\hat{y}} ~~~~~\forall(\mathbf{x},\mathbf{y})\neq (\mathbf{\hat{x}},\mathbf{\hat{y}}).
\end{eqnarray}
When only the inequality holds in the above condition, it is referred as a strict NE, otherwise the NE is a weak one~\cite{hofbauer_book,Nowak_2006_book}. A mixed (interior) NE is necessarily weak.

In the asymmetric interaction scenario, as a refinement of mixed NE, the concept of two-species ESS (2ESS)---a generalization of ESS in symmetric games---serves as the appropriate characterization of evolutionary stability. It is the states of the resident populations (1 and 2) combined such that any infinitesimal mutant fraction (in both the populations) cannot, at least be, simultaneously advantageous to the residents. Mathematically, there are three equivalent definitions~\cite{Cressman1996}:

\textbf{Definition 1a:} The state $(\mathbf{\hat{x}},\mathbf{\hat{y}}) \in \Sigma_\alpha \times \Sigma_\beta$ is a 2ESS  if there exists an $\epsilon_{(\mathbf{x},\mathbf{y})} >0$ such that, for all {non-zero} $(\epsilon_1,\epsilon_2)$  with $0 \le \epsilon_1 <\epsilon_{(\mathbf{x},\mathbf{y})} $ and $0 \le \epsilon_2 <\epsilon_{(\mathbf{x},\mathbf{y})}$,	
	\begin{eqnarray}\label{eq:(2)}
		\epsilon_1 (\mathbf{\hat{x}}-\mathbf{x})\cdot\left[ \sf{A}(\epsilon_1 \mathbf{x}+(1-\epsilon_1) \mathbf{\hat{x}})+\sf{B}(\epsilon_2 \mathbf{y}+(1-\epsilon_2) \mathbf{\hat{y}})\right] >0~~
		\text{or}~~
		\epsilon_2 (\mathbf{\hat{y}}-\mathbf{y})\cdot\left[ \sf{C}(\epsilon_2 \mathbf{x}+(1-\epsilon_1) \mathbf{\hat{x}})+\sf{D}(\epsilon_2 \mathbf{y}+(1-\epsilon_2) \mathbf{\hat{y}})\right] >0.
	\end{eqnarray}
	Here, $\epsilon_1$ and $\epsilon_2$ are the total fractions of mutants (with strategies $\mathbf{{x}}$ and $\mathbf{{y}}$) inside the  population of population 1 and population 2, respectively. $\epsilon_{(\mathbf{x},\mathbf{y})}$ is the invasion barrier for both the mutant. The definition is aptly known as the \emph{first principle definition}.

\textbf{Definition 1b:} The state $(\mathbf{\hat{x},\hat{y}})$ is a 2ESS, if there exist a neighbourhood $\mathcal{B}_{(\mathbf{\hat{x}},\mathbf{\hat{y}})} \setminus \{({\mathbf{\hat{x}}},\mathbf{\hat{y}})\}$ of  $(\mathbf{\hat{x}},\mathbf{\hat{y}})$ such that following holds:
	\begin{eqnarray}\label{eqn:neighbouhood_ESS}
		\mathbf{\hat{x}}\cdot\sf{A} \mathbf{x}+ \mathbf{\hat{x}}\cdot\sf{B} \mathbf{y}> \mathbf{x}\cdot\sf{A} \mathbf{x}+ \mathbf{x}\cdot\sf{B} \mathbf{y}
		~~~\text{or}~~
		\mathbf{\hat{y}}\cdot\sf{C} \mathbf{x}+ \mathbf{\hat{y}}\cdot\sf{D} \mathbf{y}>\mathbf{y}\cdot\sf{C} \mathbf{x}+ \mathbf{y}\cdot\sf{D} \mathbf{y}.
\end{eqnarray}
It is easy to see that 2ESS always implies NE. 

\textbf{Definition 1c:} If $(\mathbf{\hat{x},\hat{y}})$ is in interior of $\Sigma_\alpha \times \Sigma_\beta$, then $(\mathbf{\hat{x},\hat{y}})$ is a 2ESS if there is some real constant $r>0$ such that, 
	\begin{equation}
		\mathbf{\hat{x}}\cdot{\sf A}\mathbf{x}+\mathbf{\hat{x}}\cdot{\sf B} \mathbf{y}-\mathbf{x}\cdot{\sf A}\mathbf{x}-\mathbf{x}\cdot{\sf B}\mathbf{y}
		+r\left(\mathbf{\hat{y}}\cdot{\sf C}\mathbf{x}+\mathbf{\hat{y}}\cdot{\sf D} \mathbf{y}-\mathbf{y}\cdot{\sf C}\mathbf{x}-\mathbf{y}\cdot{\sf D}\mathbf{y}\right)>0,
\end{equation}
$\forall (\mathbf{x,y})\in \Sigma_\alpha \times \Sigma_\beta$ other than $(\mathbf{\hat{x},\hat{y}})$.\\
However, it is equivalent to Definitions 1a and 1b only in the case of mixed 2ESS.

In the context of symmetric games, the Folk theorems~\cite{hofbauer_book, Cressman_Tao_2014} establish interesting connections between fixed points of replicator dynamics and NE, the rational solution concepts in classical non-cooperative game theory. Furthermore, ESS has been found to be linked to the asymptotic stability of such a fixed point~\cite{taylorjonker1978,hofbauer_book,Maynard_Smith_1982,cressman_1992}. This correspondence between the game-theoretic equilibria and the fixed points of replicator dynamics assigns the idea of dynamical stability to NE and ESS. Of course, merely the existence of equilibrium outcomes is not enough: How they are dynamically achieved in the course of time is a much more practical question for evolutionary theorists.

The 2ESS is connected to dynamical stability~\cite{Cressman_Tao_2014,hofbauer_book} of fixed points in asymmetric games as well. In fact,  a non-pure 2ESS is a locally asymptotically stable rest point of the replicator equation:
\begin{subequations}
	\begin{eqnarray}
		\label{replicator_maps1}
		&&\dot{x}_i=x_i \left[({\sf A}\mathbf{x})_i+({\sf B} \mathbf{y})_i-\mathbf{x}\cdot{\sf A}\mathbf{x}-\mathbf{x}\cdot{\sf B}\mathbf{y}\right],\\\label{replicator_maps2}
		&&\dot{y}_j=y_j \left[({\sf C}\mathbf{x})_j+({\sf D}\mathbf{y})_j-\mathbf{y}\cdot{\sf C}\mathbf{x}-\mathbf{y}\cdot{\sf D}\mathbf{y}\right].
	\end{eqnarray}
\end{subequations}
Interestingly, a 2ESS in the interior of the phase space is a globally asymptotically stable rest point~\cite{Cressman_Tao_2014}.

\section{Game-theoretic equilibrium for Periodic orbits}\label{sec:III}
The question we now ask is: What about the non-convergent outcomes---like periodic orbit or chaos? How are they connected with the game-theoretic equilibria? In fact, which game-theoretic equilibria correspond to them? Investigating continuous equations to answer these questions is mathematically very challenging, and we leave it as a future exercise. 

In this paper, we employ the discrete-time version of the replicator equation---replicator maps~\cite{Pandit2018}---as the primary testbed for our quest, as we know that maps offer the most straightforward setting in which periodic solutions manifest themselves~\cite{strogatz,Vilone2011}.   Since our goal is to explore the realm of asymmetric conflicts, thus, we are motivated to explore periodic orbits in a very general setting of asymmetric conflicts where both inter and intra-specific interactions are present. 

\subsection{Time-Discrete Replicator Equation}\label{sec:tdre}
Specifically, we work with the coupled asymmetric replicator map, $\mathbf{x}^{t+1}=\mathbf{f}(\mathbf{x}^t,\mathbf{y}^t)$ and $\mathbf{y}^{t+1}=\mathbf{g}(\mathbf{x}^t,\mathbf{y}^t)$, which in component form appears as:
\begin{subequations}\label{eqn:asymmetric_replicator_gen}
	\begin{eqnarray}
		&&x^{t+1}_{i}=f_i(\mathbf{x}^t,\mathbf{y}^t)=x^t_i+x^t_i \left[({\sf A}\mathbf{x}^t)_i+({\sf B} \mathbf{y}^t)_i-\mathbf{x}^t\cdot{\sf A}\mathbf{x}^t-\mathbf{x}^t\cdot{\sf B}\mathbf{y}^t\right],\\
		&&y^{t+1}_{j}=g_i(\mathbf{x}^t,\mathbf{y}^t)=y^t_j+y^t_j \left[({\sf C}\mathbf{x}^t)_j+({\sf D}\mathbf{y}^t)_j-\mathbf{y}^t\cdot{\sf C}\mathbf{x}^t-\mathbf{y}^t\cdot{\sf D}\mathbf{y}^t\right];
	\end{eqnarray}
\end{subequations}
where $i\in\{1,2,\cdots,\alpha\}$ and $j\in\{1,2,\cdots,\beta\}$. It is straightforward to show (along the line of arguments known for bimatrix games case~\cite{Mukhopadhyay2021}) that this deterministic equation can also be interpreted as the mean-field limit of the Wright-Fisher model~\cite{wright1931,fisher1930} in a finite population. 

{\color{black} \emph{Henceforth, in this section (Section~\ref{sec:III}), for analytical tractability, we confine ourselves to the cases where each population only has two types, i.e., $\alpha=\beta=2$.}} For population 1, we denote the fraction of individuals of the first type as $x_1=x$ and hence, the fraction of individuals of the second type as $x_2=1-x$. Similarly, for population 2, we denote the fraction of individuals in the first type as $y$ and the fraction of individuals in the second type as $1-y$. Thus, the payoff matrices for population 1 can be expressed as follows:
\begin{equation}
	{\sf A} = \begin{pmatrix}
		a_{11} & a_{12} \\
		a_{21} & a_{22} 
	\end{pmatrix}
	~~~~{\sf B} = \begin{pmatrix}
		b_{11} & b_{12} \\
		b_{21} & b_{22} 
	\end{pmatrix}.
\end{equation}
Similarly, the payoff matrices of population 2 are, 
\begin{equation}
	{\sf C}=\begin{pmatrix}
		c_{11} & c_{12} \\
		c_{21} & c_{22} 
	\end{pmatrix}
	~~~~{\sf D}=\begin{pmatrix}
		d_{11} & d_{12} \\
		d_{21} & d_{22} 
	\end{pmatrix}.
\end{equation}
Eq.~(\ref{eqn:asymmetric_replicator_gen}) takes the following form~\cite{Schuster_1981_III}:
\begin{subequations} \label{eqn:replicator2by2}
	\begin{eqnarray}
		&&x^{t+1}=f(x^t,y^t)\equiv x^t+x^t(1-x^t)(a+bx^t+cy^t),\\
		&&y^{t+1}=g(x^t,y^t)\equiv y^t+y^t(1-y^t)(d+rx^t+sy^t),
	\end{eqnarray}
\end{subequations}
where, $a\equiv a_{12}+b_{12}-a_{22}-b_{22}$,~$b\equiv a_{11}-a_{12}-a_{21}+a_{22}$,~ $c\equiv b_{11}-b_{12}-b_{21}+b_{22}$,~
~$d\equiv c_{12}+d_{12}-c_{22}-d_{22}$,~$r\equiv c_{11}-c_{12}-c_{21}+c_{22}$, and $s\equiv d_{11}-d_{12}-d_{21}+d_{22}$. The forward invariance of the phase space $\Sigma_2\times\Sigma_2$ is guaranteed only if particular combinations of values of the parameters are chosen. 

In other words, arbitrary values for the payoff parameters $a,\,b,\,c,\,d,\,r$ and $s$ in Eq.~(\ref{eqn:replicator2by2}), the map may yield a mathematical solution tracing a trajectory in the phase space $\mathbb{R}\times\mathbb{R}$ such that the trajectory might not be bounded in the phase space: i.e., $(\mathbf{x}^{t+1},\mathbf{y}^{t+1})\notin\Sigma_{2}\times\Sigma_2$ even though $(\mathbf{x}^t,\mathbf{y}^t)\in\Sigma_{2}\times \Sigma_{2}$ at some time $t$. Appositely, the replicator map is said to possesses \emph{physical solutions} if they consistently map $\Sigma_{2}\times \Sigma_{2}\to \Sigma_{2}\times \Sigma_{2}$ for all $(\mathbf{x}^t,\mathbf{y}^t)\in\Sigma_{2}\times \Sigma_{2}$ at all times, $t$ (see Ref.~\cite{Pandit2018,MCC_JPC_21}). The region in the parameter space (in the present case, a six-dimensional one) where the map has physical solutions is called \emph{strict physical region}. 

In effect, our objective is to find the conditions on the parameters such that $\forall (x,y)\in[0,1]\times[0,1]$ both $f(x,y)$ and $g(x,y)$ remains bounded within $[0,1]$ simultaneously. We first present arguments for $f(x,y)$ as the arguments straightforwardly can be extended for $g(x,y)$. First, we note that $f(x,y)$ is a cubic  in $x$ and linear in $y$---see Eq.~(\ref{eqn:replicator2by2}). Thus, all we have to do is to find the set of parameter values for which at $y=0$ and $y=1$ the function remains between zero to one for all $x$-values; because of linearity in $y$, $f(x,y)$ automatically remains within $[0,1]$ $\forall (x,y)$.

At $y=0$ and $y=1$, the forms of $f(x,y)$ are $f(x;\chi)=x+x(1-x)(a +bx+\chi c)$ with $\chi=0$ and $\chi=1$, respectively. Note that $f(0;\chi)=0$ and $f(1;\chi)=1$. Also, note that $f(x;\chi)$---being a cubic polynomial---must have at most two extrema. Consequently, 
\begin{enumerate} 
	\item $f(x;\chi)$ remains inside $[0,1]$ irrespective of parameter values if no extremum is inside $(0,1)$; 
	\item $f(x;\chi)$ may remain inside $[0,1]$ for a certain set of parameter values if all the extrema are inside $[0,1]$. The appropriate parameter values should be chosen such that the extrema, $f(x^\star_-;\chi)$ and $f(x^\star_+;\chi)$, evaluated at 
	\begin{equation}x^\star_{\pm}=\frac{b-a-\chi c}{3b}\pm\sqrt{\left(\frac{b-a-\chi c}{3b}\right)^2+\frac{1+a+\chi c}{3b}},
	\end{equation}
	lie between zero and one.
\end{enumerate}
It is apparent that exactly analogous conditions can be given for $g(x,y)$: All one has to do is replace $x$, $a$, $b$ and $c$ by $y$, $d$, $s$ and $r$, respectively, in the immediately preceding conditions. 

Having presented the mechanism for finding the strict physical region, we defer its numerical verification to Sec.~\ref{sec:numver} where the existence of periodic orbits, which remain bounded within the allowed phase space, is depicted as well.
\subsection{Extension of Nash Equilibrium: Heterogeneity Orbit}\label{sec:3b}
Now, let us extend the definitions of game-theoretic equilibria such that the extensions can be related to the periodic outcomes. The discrete replicator map for the asymmetric game, which maps the frequency of phenotype in the $k$-th generation to the frequency of phenotype in the $(k+1)$-th generation, can be recast based on Eq.~(\ref{eqn:asymmetric_replicator_gen}) as follows:
\begin{subequations}\label{eqn:replicator2}
\begin{eqnarray}
	x^{k+1}=x^k+x^k(1-x^k)[(\sf{A}\mathbf{x}^k)_1+(\sf{B}\mathbf{y}^k)_1-\sf({A}\mathbf{x}^k)_2-(\sf{B}\mathbf{y}^k)_2],\label{bimatrix_replicator_map1}\\
	y^{k+1}=y^k+y^k(1-y^k)[(\sf{C}\mathbf{x}^k)_1+(\sf{D}\mathbf{y}^k)_1-\sf({C}\mathbf{x}^k)_2-(\sf{D}\mathbf{y}^k)_2].\label{bimatrix_replicator_map2}
\end{eqnarray}
\end{subequations}
Let a sequence of states given by $\{(\mathbf{\hat{x}}^k,\mathbf{\hat{y}}^k): (\hat{x}^k,\hat{y}^k)\in (0,1)\times(0,1),~ k=1,2,\cdots,m \}$ with $(\mathbf{\hat{x}}^i,\mathbf{\hat{y}}^i) \ne (\mathbf{\hat{x}}^j,\mathbf{\hat{y}}^j)$  $\forall i \ne j$, represents a periodic orbit with period $m$ such that, $(\mathbf{\hat{x}}^{m+1},\mathbf{\hat{y}}^{m+1})=(\mathbf{\hat{x}}^1,\mathbf{\hat{y}}^1)$. Now, summing all the $m$ generations of Eq.~(\ref{eqn:replicator2}), we get,
\begin{subequations}\label{eqn:periodicity_condition}
	\begin{eqnarray}
		&& \sum_{k=1}^{m}H_{\mathbf{\hat{x}}^k}\left[(\sf{A}\mathbf{\hat{x}}^k)_1+(\sf{B}\mathbf{\hat{y}}^k)_1-(\sf{A}\mathbf{\hat{x}}^k)_2-(\sf{B}\mathbf{\hat{y}}^k)_2~\right]=0,\\
		&&\sum_{k=1}^{m}H_{\mathbf{\hat{y}}^k}\left[(\sf{C}\mathbf{\hat{x}}^k)_1+(\sf{D}\mathbf{\hat{y}}^k)_1-(\sf{C}\mathbf{\hat{x}}^k)_2-(\sf{D}\mathbf{\hat{y}}^k)_2 ~\right]=0.
	\end{eqnarray}
\end{subequations}
Here, we define, heterogeneity, $H_{\mathbf{x}^k}\equiv2x^k (1-x^k)$, which is the probability that two randomly chosen individuals of the corresponding population are of two different types. It quantifies how heterogeneous-strategied the population is. Furthermore, we can define heterogeneity weighted payoffs~\cite{archan_periodic}, i.e., payoffs multiplied with respective heterogeneity factors $H$. The Nash equilibrium condition Eq.~(\ref{11}) can be, rather trivially, rewritten for the mixed NE in terms of heterogeneity weighted payoffs as,
\begin{subequations}\label{eqn:HO_fixed_point}
	\begin{eqnarray}
		&&H_{\mathbf{\hat{x}}}	\left[\mathbf{\hat{x}}\cdot\sf{A} \mathbf{\hat{x}}+\mathbf{\hat{x}}\cdot\sf{B} \mathbf{\hat{y}}\right] = H_{\mathbf{\hat{x}}}\left[\mathbf{{x}}\cdot\sf{A} \mathbf{\hat{x}}+\mathbf{{x}}\cdot\sf{B} \mathbf{\hat{y}}\right] ~~~~\forall\mathbf{x} \in \Sigma_{\alpha}, \\
		&&H_{\mathbf{\hat{y}}}	\left[\mathbf{\hat{y}}\cdot\sf{C} \mathbf{\hat{x}}+\mathbf{\hat{y}}\cdot\sf{D} \mathbf{\hat{y}}\right] = H_{\mathbf{\hat{y}}}\left[\mathbf{{y}}\cdot\sf{C} \mathbf{\hat{x}}+\mathbf{{y}}\cdot\sf{D} \mathbf{\hat{y}}\right]~~~~~\forall \mathbf{y} \in \Sigma_\beta.
	\end{eqnarray}
\end{subequations}
This condition is obtained from Eqs.~(\ref{eqn:periodicity_condition}) in the special case of a 1-period orbit or a fixed point. Thus, it is a natural motivation to adopt the concept of heterogeneity weighted orbit, HO($m$), as an extension of Nash equilibrium for an $m$-periodic orbit:

\textbf{Definition 2:} The sequence of distinct states $\{(\mathbf{\hat{x}}^k,\mathbf{\hat{y}}^k):(\hat{x}^k,\hat{y}^k) \in (0,1)\times(0,1),~ k=1,2,\cdots,m \}$ (with $(\mathbf{\hat{x}}^i,\mathbf{\hat{y}}^i)\ne(\mathbf{\hat{x}}^j,\mathbf{\hat{y}}^j)~\forall~i \ne j$) is termed as HO($m$) if it satisfies the following two equalities $\forall x\in(0,1)$ and $y\in(0,1)$:
\begin{subequations} \label{eqn:HO_definition}
	\begin{eqnarray}
		&&\sum_{k=1}^{m} H_{\mathbf{\hat{x}}^k} \left[\mathbf{\hat{x}}^k\cdot\sf{A} \mathbf{\hat{x}}^k+\mathbf{\hat{x}}^k\cdot\sf{B} \mathbf{\hat{y}}^k\right] = \sum_{k=1}^{m} H_{\mathbf{\hat{x}}^k}\left[\mathbf{{x}}\cdot\sf{A} \mathbf{\hat{x}}^k+\mathbf{{x}}\cdot\sf{B} \mathbf{\hat{y}}^k\right],~~~~\\  
		&& \sum_{k=1}^{m} H_{\mathbf{\hat{y}}^k}\left[\mathbf{\hat{y}}^k\cdot\sf{C} \mathbf{\hat{x}}^k+\mathbf{\hat{y}}^k\cdot\sf{D} \mathbf{\hat{y}}^k\right] =\sum_{k=1}^{m} H_{\mathbf{\hat{y}}^k}\left[\mathbf{{y}}\cdot\sf{C} \mathbf{\hat{x}}^k+\mathbf{{y}}\cdot\sf{D} \mathbf{\hat{y}}^k\right].
	\end{eqnarray}
\end{subequations}
 If Eq.~(\ref{eqn:HO_definition}) holds then,
\begin{subequations} \label{20}
	\begin{eqnarray}
		&& \frac{d}{dx}\sum_{k=1}^{m} H_{\mathbf{\hat{x}}^k}\left[\mathbf{{x}}\cdot\sf{A} \mathbf{\hat{x}}^k+\mathbf{{x}}\cdot\sf{B} \mathbf{\hat{y}}^k\right]=0, \\
		&&\frac{d}{dy}\sum_{k=1}^{m} H_{\mathbf{\hat{y}}^k}\left[\mathbf{{y}}\cdot\sf{C} \mathbf{\hat{x}}^k+\mathbf{{y}}\cdot\sf{D} \mathbf{\hat{y}}^k\right] =0,
	\end{eqnarray}
\end{subequations}
which in turn implies:
\begin{subequations}\label{21}
	\begin{eqnarray}
		&&\sum_{k=1}^{m} H_{\mathbf{\hat{x}}^k}\left[(\sf{A}\mathbf{\hat{x}}^k)_1+(\sf{B}\mathbf{\hat{y}}^k)_1-(\sf{A}\mathbf{\hat{x}}^k)_2-(\sf{B}\mathbf{\hat{y}}^k)_2\right]=0, \\ 
		&&\sum_{k=1}^{m} H_{\mathbf{\hat{y}}^k}\left[ (\sf{C}\mathbf{\hat{x}}^k)_1+(\sf{D}\mathbf{\hat{y}}^k)_1-(\sf{C}\mathbf{\hat{x}}^k)_2-(\sf{D}\mathbf{\hat{y}}^k)_2\right]=0.
	\end{eqnarray}
\end{subequations}
This matches with Eq.~(\ref{eqn:periodicity_condition}). We can easily see that one trivial possibility is the square bracket terms in Eq.~(\ref{21}) are individually zero, which corresponds to NE. But another non-trivial possibility is that the overall sum is zero; whether such a solution exists depends on the exact structure of payoff matrices. It is this non-trivial solution that is a possible candidate for HO($m$) relating to the non-convergent outcomes.

\subsection{Extension of two-species ESS: Heterogeneity Stable Orbit}\label{sec:3c}
The concept of a two-species ESS states that neither of the two mutant states can successfully invade host populations~\cite{cressman_1992,Cressman1996,Cressman_Tao_2014}. Similar to a single-species ESS, a (interior) two-species ESS is locally asymptotically stable under replicator dynamics. In this section, our objective is to explore the extension of a two-species ESS, which corresponds to a locally asymptotically stable periodic orbit in the discrete replicator dynamics. It is evident that we need to concentrate on Definition 1b of a two-species ESS. We name the extension: two-species HSO (2HSO), which should be a refinement of HO($m$) since ESS is the refinement NE.

\textbf{Definition 3:} 2HSO($m$) of the two-dimensional map---$x^{k+1}=f(\mathbf{x}^k,\mathbf{y}^k)$~and~$y^{k+1}=g(\mathbf{x}^k,\mathbf{y}^k)$---is a sequence of states $\{(\mathbf{\hat{x}}^k,\mathbf{\hat{y}}^k):(\hat{x}^k,\hat{y}^k) \in (0,1)\times(0,1),~ k=1,2, \cdots,m;~ (\mathbf{\hat{x}}^{i},\mathbf{\hat{y}}^{i})\neq(\mathbf{\hat{x}}^{j},\mathbf{\hat{y}}^{j}) ~\forall i \neq j \}$ such that:
\begin{subequations}\label{eqn:HSO}
	\begin{eqnarray}
	&&\sum_{k=1}^{m} H_{\mathbf{{x}}^k}\left[\mathbf{\hat{x}}^1\cdot\sf{A}\mathbf{x}^k+\mathbf{\hat{x}}^1\cdot \sf{B}\mathbf{y}^k \right] > 	\sum_{k=1}^{m} H_{\mathbf{{x}}^k} \left[\mathbf{{x}}^1\cdot\sf{A} \mathbf{{x}}^k+\mathbf{{x}}^1\cdot\sf{B} \mathbf{{y}}^k\right],\\   \nonumber &&\text{or}, \\
	&&\sum_{k=1}^{m} H_{\mathbf{{y}}^k} \left[\mathbf{\hat{y}}^1\cdot\sf{C} \mathbf{{x}}^k+\mathbf{\hat{y}}^1\cdot\sf{D} \mathbf{y}^k \right] >\sum_{k=1}^{m} H_{\mathbf{{y}}^k}  \left[\mathbf{{y}}^1\cdot\sf{C} \mathbf{{x}}^k+\mathbf{{y}}^1\cdot\sf{D} \mathbf{{y}}^k \right],
	\end{eqnarray}
\end{subequations}
for any orbit $\left\{(\mathbf{{x}}^k,\mathbf{{y}}^k)\right\}_{k=1}^{k=m}$ of the map starting at an infinitesimal (deleted) neighbourhood $\mathcal{B}_{(\mathbf{\hat{x}}^1,\mathbf{\hat{y}}^1)} \setminus \{({\mathbf{\hat{x}}}^1,\mathbf{\hat{y}}^1)\}$ of  $(\mathbf{\hat{x}}^1,\mathbf{\hat{y}}^1)$.

 In game-theoretical language, we can say that 2HSO($m$) is a strategy profile of length $m$ such that when a resident population adopts this, then there exists no rare mutant adopting a close but different strategy profile which can invade the resident population. We can directly check that for $m=1$ Eq.~(\ref{eqn:HSO}) matches with Eq.~(\ref{eqn:neighbouhood_ESS}), i.e., the definition of 2ESS.

We now prove that 2HSO($m$) is a refinement of HO($m$) that is, 2HSO implies HO. The inequalities in Eq.~(\ref{eqn:HSO}) can be recast as:
\begin{subequations}\label{eqn:HSO_step}
	\begin{eqnarray}
		&&(x^1-\hat{x}^1)\sum_{k=1}^{m} H_{\mathbf{{x}}^k} \left[(\sf{A}\mathbf{x}^k)_1-({\sf A}\mathbf{x}^k)_2+({\sf B}\mathbf{y}^k)_1-({\sf B}\mathbf{y}^k)_2\right]<0,\\
		\text{or}, \nonumber &&\\
		&&(y^1-\hat{y}^1)	\sum_{k=1}^{m}
		H_{\mathbf{{y}}^k} \left[({\sf C}\mathbf{x}^k)_1-({\sf C}\mathbf{x}^k)_2+({\sf D}\mathbf{y}^k)_1-({\sf D}\mathbf{y}^k)_2 \right]<0.
	\end{eqnarray}
\end{subequations}
Since the sign of the two terms have to be opposite for the product to be negative for all $(x,y)$ near $(\hat{x},\hat{y})$, the following is true,
\begin{subequations}\label{eqn:HSO_step2}
	\begin{eqnarray}
		 &&\lim\limits_{(x^1-\hat{x}^1) \to 0} \lim\limits_{(y^1-\hat{y}^1) \to 0}\sum_{k=1}^{m} H_{\mathbf{{x}}^k} \left[({\sf A}\mathbf{x}^k)_1-(A\mathbf{x}^k)_2 +({\sf B}\mathbf{y}^k)_1-({\sf B}\mathbf{y}^k)_2 \right]=0,\\ \nonumber && \text{and} \\  &&   \lim\limits_{(x^1-\hat{x}^1) \to 0}\lim\limits_{(y^1-\hat{y}^1) \to 0} 	\sum_{k=1}^{m} 
		H_{\mathbf{{y}}^k} \left[({\sf C}\mathbf{x}^k)_1-({\sf C}\mathbf{x}^k)_2 +({\sf D}\mathbf{y}^k)_1-({\sf D}\mathbf{y}^k)_2  \right]=0. 
	\end{eqnarray}
\end{subequations}
Notice that the change in the change from "or" to "and" from Eq.~(\ref{eqn:HSO_step}) to Eq.~(\ref{eqn:HSO_step2}).  This is because if both the inequalities are satisfied in Eq.~(\ref{eqn:HSO_step}), then, Eq.~(\ref{eqn:HSO_step2}) is satisfied. But even if one of the inequalities is satisfied in  Eq.~(\ref{eqn:HSO_step}) then also it implies Eq.~(\ref{eqn:HSO_step2}); this is because the sign of terms have to be same for the inequality to be greater than zero, so it crosses zero at the limit, $\lim{(x^1-\hat{x}^1)\to0}\lim{(y^1-\hat{y}^1)\to0}$. Finally we get,
\begin{subequations}\label{43}
	\begin{eqnarray}
		&& \sum_{k=1}^{m} H_{\mathbf{\hat{x}}^k} \left[({\sf A}\mathbf{\hat{x}}^k)_1-({\sf A}\mathbf{\hat{x}}^k)_2+({\sf B}\mathbf{\hat{y}}^k)_1-({\sf B}\mathbf{\hat{y}}^k)_2\right]=0,\\\nonumber && \text{and} \\  &&  \sum_{k=1}^{m} 		H_{\mathbf{\hat{y}}^k} \left[({\sf C}\mathbf{\hat{x}}^k)_1-({\sf C}\mathbf{\hat{x}}^k)_2+({\sf D}\mathbf{\hat{y}}^k)_1-({\sf D}\mathbf{\hat{y}}^k)_2 \right]=0. 
	\end{eqnarray}
\end{subequations}
Comparing Eq.~(\ref{43}) and Eq.~(\ref{21}), we conclude that 2HSO($m$) implies HO($m$).
\subsection{HSO and dynamical stability}\label{sec:3E}
We know NE corresponds to a fixed point of replicator dynamics, and the condition of ESS is related to the asymptotic stability of the fixed point. Extending NE, we reach HO($m$). Periodic orbit in discrete replicator dynamics corresponds to HO($m$). Also, extending 2ESS, we define 2HSO($m$). So it is natural to guess that the stability of periodic orbit may have some connection with 2HSO($m$)---this is what we are going to show below.

\textbf{Proposition 1:} If the sequence of states $\{(\mathbf{\hat{x}}^k,\mathbf{\hat{y}}^k): (\hat{x}^k,\hat{y}^k) \in(0,1)\times(0,1);~ k=1,2,\cdots,m\}$, where $(\mathbf{\hat{x}}^i,\mathbf{\hat{y}}^i) \ne (\mathbf{\hat{x}}^j,\mathbf{\hat{y}}^j)$, $\forall i\ne j$, is a locally asymptotically stable $m$-period orbit of the replicator map Eq.~(\ref{eqn:replicator2}) for asymmetric game then it must be 2HSO($m$).

\emph{Proof:} Let the sequence of distinct states $\{(\mathbf{\hat{x}}^1,\mathbf{\hat{y}}^1),(\mathbf{\hat{x}}^2,\mathbf{\hat{y}}^2),\cdots,(\mathbf{\hat{x}}^m,\mathbf{\hat{y}}^m)\}$ be a locally asymptotically stable periodic orbit of the replicator map. Then for some state, $(\mathbf{x}^1,\mathbf{y}^1)$, in the neighbourhood of $(\mathbf{\hat{x}}^1,\mathbf{\hat{y}}^1)$, we can write,
\begin{equation}\label{eqn:HSO_stability}
	\frac{\left[\left\{f^m (\mathbf{x}^1,~\mathbf{y}^1)-\hat{x}^1 \right\}^2 + \left\{g^m (\mathbf{x}^1,~\mathbf{y}^1)-\hat{y}^1 \right\}^2\right]^{1/2}}{\left[(x^1-\hat{x}^1)^2 + (y^1-\hat{y}^1)^2\right]^{1/2}} <1.
\end{equation}
Using the explicit form of $f^m (\mathbf{x}^1,~\mathbf{y}^1)$ and $g^m (\mathbf{x}^1,~\mathbf{y}^1)$ we get,
\begin{eqnarray}\label{eqn:HSO_stability1}
 \frac{1}{\left[(x^1-\hat{x}^1)^2 + (y^1-\hat{y}^1)^2\right]^{1/2}}	\Bigg[\Big\{(x^1-\hat{x}^1) + \frac{1}{2} \sum_{k=1}^{m} H_{\mathbf{x}^k} \left[(\sf{A}\mathbf{x}^k)_1-(\sf{A}\mathbf{{x}}^k)_2+(\sf{B}\mathbf{y}^k)_1-(\sf{B}\mathbf{{y}}^k)_2 \right]\Big\}^2\nonumber\\
 + \Big\{(y^1-\hat{y}^1) + \frac{1}{2} \sum_{k=1}^{m} 
		H_{\mathbf{{y}}^k}\left[(\sf{C}\mathbf{{x}}^k)_1-(\sf{C}\mathbf{{x}}^k)_2+(\sf{D}\mathbf{{y}}^k)_1-(\sf{D}\mathbf{{y}}^k)_2 \right]\Big\}^2\Bigg]^{1/2} <1.
\end{eqnarray}
It follows that it is necessary for the stability of the periodic orbit, one of the following three cases hold in the LHS of Eq.~(\ref{eqn:HSO_stability1}):
\begin{enumerate}
\item If signs of the two terms inside the first square term in the numerator are same, then the signs of the two terms inside the second square term must be different---making the numerator less than the denominator. Mathematically,
	if, $(x^1-\hat{x}^1)	\sum_{k=1}^{m} H_{\mathbf{{x}}^k} \left[(\sf{A}\mathbf{{x}}^k)_1-(\sf{A}\mathbf{{x}}^k)_2+(\sf{B}\mathbf{y}^k)_1-(\sf{B}\mathbf{{y}}^k)_2\right]>0$
	~then~
	$(y-\hat{y}^1)\sum_{k=1}^{m}
	H_{\mathbf{{y}}^k}.\left[(\sf{C}\mathbf{{x}}^k)_1-(\sf{C}\mathbf{{x}}^k)_2+(\sf{D}\mathbf{{y}}^k)_1-(\sf{D}\mathbf{{y}}^k)_2\right]<0$.
\item Similarly, if the signs of the two terms inside the second square term in the numerator are same, then the signs of the two terms inside the first square term must be different. Mathematically,
if,	$(y^1-\hat{y}^1)\sum_{k=1}^{m}
	H_{\mathbf{{y}}^k}\left[(\sf{C}\mathbf{{x}}^k)_1-(\sf{C}\mathbf{{x}}^k)_2+(\sf{D}\mathbf{{y}}^k)_1-(\sf{D}\mathbf{{y}}^k)_2\right]>0$ ~then~
	$(x^1-\hat{x}^1)	\sum_{k=1}^{m} H_{\mathbf{{x}}^k} \left[(\sf{A}\mathbf{{x}}^k)_1-(\sf{A}\mathbf{{x}}^k)_2+(\sf{B}\mathbf{y}^k)_1-(\sf{B}\mathbf{{y}}^k)_2 \right]<0$.
\item Lastly, if signs of the two terms inside each square term in the numerator are different. Mathematically,
	$(x^1-\hat{x}^1)	\sum_{k=1}^{m} H_{\mathbf{{x}}^k} \left[(\sf{A}\mathbf{{x}}^k)_1-(\sf{A}\mathbf{{x}}^k)_2+(\sf{B}\mathbf{y}^k)_1-(\sf{B}\mathbf{{y}}^k)_2 \right]<0$~and~
	$(y^1-\hat{y}^1)\sum_{k=1}^{m}
	H_{\mathbf{{y}}^k}\left[(\sf{C}\mathbf{{x}}^k)_1-(\sf{C}\mathbf{{x}}^k)_2+(\sf{D}\mathbf{{y}}^k)_1-(\sf{D}\mathbf{{y}}^k)_2\right]<0.$
\end{enumerate}

Following arguments akin to the one presented while going from Eq.~(\ref{eqn:HSO}) to Eq.~(\ref{eqn:HSO_step}), these three conditions can be written simultaneously as,
\begin{subequations}\label{29}
	\begin{eqnarray}
		&&\sum_{k=1}^{m} H_{\mathbf{{x}}^k}\left[ \mathbf{\hat{x}}^1\cdot\sf{A} \mathbf{{x}}^k +\mathbf{\hat{x}}^1\cdot\sf{B} \mathbf{{y}}^k\right] > 	\sum_{k=1}^{m}  H_{\mathbf{{x}}^k} \left[\mathbf{{x}}^1\cdot\sf{A} \mathbf{{x}}^k+\mathbf{{x}}^1\cdot\sf{B} \mathbf{{y}}^k \right],\\ \nonumber && \text{or}, \\
		&&\sum_{k=1}^{m} H_{\mathbf{{y}}^k} \left[\mathbf{\hat{y}}^1\cdot\sf{C} \mathbf{{x}}^k+\mathbf{\hat{y}}^1\cdot\sf{D} \mathbf{{y}}^k \right] >\sum_{k=1}^{m} H_{\mathbf{{y}}^k}  \left[\mathbf{{y}}^1\cdot\sf{C} \mathbf{{x}}^k+\mathbf{{y}}^1\cdot\sf{D} \mathbf{{y}}^k \right].
	\end{eqnarray}
\end{subequations}  		
This is nothing but the 2HSO($m$)'s definition (see Definition 3). This completes the proof.

The converse of the theorem---2HSO($m$) implies locally asymptotically stable ${m}$ period orbit---is not always true. For example, we can see that even if condition (1) is true, it is possible to have $\big\vert x^1-\hat{x}^1\big\vert\big\vert\sum_{k=1}^{m} H_{\mathbf{{x}}^k} \left[(\sf{A}\mathbf{{x}}^k)_1-(\sf{A}\mathbf{{x}}^k)_2+(\sf{B}\mathbf{y}^k)_1-(\sf{B}\mathbf{{y}}^k)_2 \right]\big\vert >$ $\big\vert y^1-\hat{y}^1\big\vert \big\vert\sum_{k=1}^{m} H_{\mathbf{{x}}^k} \left[(\sf{C}\mathbf{{x}}^k)_1-(\sf{C}\mathbf{{x}}^k)_2+(\sf{D}\mathbf{y}^k)_1-(\sf{D}\mathbf{{y}}^k)_2 \right]\big\vert$, such that the periodic orbit is unstable as inequality (\ref{eqn:HSO_stability1}) may not be satisfied. This is because if we expand the square terms, the positive terms dominate the negative terms rendering the numerator greater than the denominator. Thus, for the converse to hold true, additional conditions are required. Specifically,  letting~$E\equiv\frac{1}{2} \sum_{k=1}^{m} H_{\mathbf{x}^k} \left[(\sf{A}\mathbf{x}^k)_1-(\sf{A}\mathbf{x}^k)_2+(\sf{B}\mathbf{y}^k)_1-(\sf{B}\mathbf{y}^k)_2 \right]$ and
$F\equiv\frac{1}{2}\sum_{k=1}^{m}
H_{\mathbf{\hat{y}}^k}\left[(\sf{C}\mathbf{x}^k)_1-(\sf{C}\mathbf{x}^k)_2+(\sf{D}\mathbf{y}^k)_1-(\sf{D}\mathbf{y}^k)_2\right]$, the additional condition corresponding to condition (1) is
\begin{equation}
-2F(y^1-\hat{y}^1)-F^2 > E \left[2(x^1-\hat{x}^1)+E \right],
\end{equation}
corresponding to condition (2) is
\begin{equation}
-2E (x^1-\hat{x}^1)-E^2 > F \left[2(y^1-\hat{y}^1)+F \right],
\end{equation}
and corresponding to condition (3) is
\begin{equation}
	-2E (x^1-\hat{x}^1)-E^2  > F \left[F+2(y^1-\hat{y}^1) \right].
\end{equation}
\subsection{Connection with Strong Stability}
It is well-known that the connection between the concept of strong stability~\cite{hofbauer_book} and evolutionarily stable strategy is paramount and gives validation to the idea of evolutionary stability. Specifically, a strategy is evolutionary stable if and only if it is strongly stable, which indicates that any convex combination of types (i.e., mean population strategy) converges to the strategy. Let us develop here the extension of aforementioned connection to the case of oscillatory outcomes.

Recalling the notations discussed in at the beginning of Sec.~\ref{sec:convergent_outcomes}, let $(\mathbf{p}_i,\mathbf{p}'_i)$ be the strategies of the underlying game for population 1 and $(\mathbf{q}'_i,\mathbf{q}_i)$ be the strategy of underlying game for population 2. We define $(\mathbf{\bar{p}}, \mathbf{\bar{p}'})=(\sum_ix_i\mathbf{p}_i,\sum_ix_i\mathbf{p}'_i)$ as the average population strategy of population 1; similarly, we define $(\mathbf{\bar{q}'},\mathbf{\bar{q}} )=(\sum_iy_i\mathbf{q}'_i,\sum_iy_i\mathbf{q}_i)$ as the average population strategy of population 2. For simplicity and clearer presentation, we confine ourselves to the case where both the populations have only two types with frequencies denoted by (${x},1-{x}$) and (${y},1-{y}$). We propose the following definition.

\textbf{Definition 4:} A sequence of strategies profiles
	 $\{\left((\mathbf{\hat{\bar{p}}}^k,\mathbf{\hat{\bar{p}}}'^k),(\mathbf{\hat{\bar{q}}}'^k,\mathbf{\hat{\bar{q}}}^k)\right):\mathbf{\hat{\bar{p}}}^k=\sum_{i=1}^2\hat{x}_i^k\mathbf{p}_i,~\mathbf{\hat{\bar{p}}}'^k=\sum_{i=1}^2\hat{x}_i^k\mathbf{p}'_i,~\mathbf{\hat{\bar{q}}}'^k=\sum_{i=1}^2\hat{y}_i^k\mathbf{q}'_i~\textrm{and}~\mathbf{\hat{\bar{q}}}^k=\sum_{i=1}^2\hat{y}_i^k\mathbf{q}_i ;~(\hat{x}^k,\hat{y}^k)\in(0,1)\times(0,1) \forall k=1,2,\cdots,m;~\mathbf{p}_i\in\Sigma_{N_{11}},~\mathbf{p}'_i\in\Sigma_{N_{12}},~\mathbf{q}_i\in\Sigma_{N_{21}}~\textrm{and}~\mathbf{q}'_i\in\Sigma_{N_{22}} \}$ where $\left((\mathbf{\hat{\bar{p}}}^i,\mathbf{\hat{\bar{p}}}'^i),(\mathbf{\hat{\bar{q}}}'^i,\mathbf{\hat{\bar{q}}}^i)\right)\neq\left((\mathbf{\hat{\bar{p}}}^j,\mathbf{\hat{\bar{p}}}'^j),(\mathbf{\hat{\bar{q}}}'^j,\mathbf{\hat{\bar{q}}}^j)\right)$ $\forall i\neq j$ is an SSSS($m$) (strongly stable strategy set) if any initial strategy profile $\left(({\mathbf{\bar{p}}}^i,{\mathbf{\bar{p}}}'^i),({\mathbf{\bar{q}}}'^i,{\mathbf{\bar{q}}}^i)\right)$ that is sufficiently close to SSSS, converges to SSSS.\\
\textbf{Proposition 2:} If $\left\{(\mathbf{\hat{\bar{p}}}^k,\mathbf{\hat{\bar{p}}}'^k),(\mathbf{\hat{\bar{q}}}'^k,\mathbf{\hat{\bar{q}}}^k)\right\}_{k=1}^{k=m}$ is SSSS($m$), then $\{(\mathbf{\hat{x}}^k,\mathbf{\hat{y}}^k)\}_{k=1}^{k=m}$ is 2HSO($m$).

\emph{Proof:} By definition $\mathbf{\hat{\bar{p}}}^k=\hat{x}^k\mathbf{p}_1+(1-\hat{x}^k)\mathbf{p}_2$,~$\mathbf{\hat{\bar{p}}}'^k=\hat{x}^k\mathbf{p}'_1+(1-\hat{x}^k)\mathbf{p}'_2$,~$\mathbf{\hat{\bar{q}}}'^k=\hat{y}^k\mathbf{q}'_1+(1-\hat{y}^k)\mathbf{q}'_2$~and~$\mathbf{\hat{\bar{q}}}^k=\hat{y}^k\mathbf{q}_1+(1-\hat{y}^k)\mathbf{q}_2$. Any infinitesimal perturbation about an element of SSSS can be represented as $\mathbf{\hat{\bar{p}}}^k+\epsilon(\mathbf{p}_1-\mathbf{p}_2)=(\hat{x}^k+\epsilon)\mathbf{p}_1+(1-\hat{x}^k-\epsilon)\mathbf{p}_2$,~$\mathbf{\hat{\bar{p}}}'^k+\epsilon(\mathbf{p}'_1-\mathbf{p}'_2)=(\hat{x}^k+\epsilon)\mathbf{p}'_1+(1-\hat{x}^k-\epsilon)\mathbf{p}'_2$,~$\mathbf{\hat{\bar{q}}}'^k+\epsilon(\mathbf{q}'_1-\mathbf{q}'_2)=(\hat{y}^k+\epsilon)\mathbf{q}'_1+(1-\hat{y}^k-\epsilon)\mathbf{q}'_2$~and~$\mathbf{\hat{\bar{q}}}^k+\epsilon(\mathbf{q}_1-\mathbf{q}_2)=(\hat{y}^k+\epsilon)\mathbf{q}_1+(1-\hat{y}^k-\epsilon)\mathbf{q}_2$ where $|\epsilon|$ is sufficiently small. Thus, we note that if any initial average population strategy is sufficiently close to an element of SSSS, then in the population dynamics, the initial state is sufficiently close to the corresponding element of the sequence of states $\{(\mathbf{\hat{x}}^k,\mathbf{\hat{y}}^k):(\hat{x}^k,\hat{y}^k)\in(0,1)\times(0,1),k=1,2,\cdots,m\}$. Since, by  definition, the initial average population strategy converges to SSSS($m$), if we start sufficiently close to any state of the sequence $\{(\mathbf{\hat{x}}^k,\mathbf{\hat{y}}^k):(\hat{x}^k,\hat{y}^k)\in(0,1)\times(0,1), k=1,2,\cdots,m\}$, then the population state must converge to this set. Hence, the set of states is locally asymptotically stable $m$-periodic orbit and therefore, it must be 2HSO$(m)$ in line with proposition 1 proven in Section~\ref{sec:3E}.

We remark that the converse of proposition 2 does not always hold good as a 2HSO$(m)$ need not be locally asymptotically stable $m$-periodic orbit.
\section{Information theoretic interpretation}\label{sec:info}
In this section, we adopt an information-theoretic perspective to guide us in interpreting periodic orbits. We find that the relative entropy, also known as the Kullback--Leibler (KL) divergence~\cite{kullback1951,cover1999}, to be especially valuable for this purpose. It is defined as follows,
\begin{equation}\label{eqn:KL_divergence}
	D_{KL}(\mathbf{p}||\mathbf{q})=\sum_ip_i\log\left[\frac{ p_i}{q_i}\right],
\end{equation}
where $i$ belongs to the support of probability distributions
$\mathbf{p}$ and $\mathbf{q}$ for which the KL-divergence is defined. The KL-divergence serves as a valuable tool for quantifying the dissimilarity between two probability distributions. The KL-divergence can obviously be defined using any two states of a population as any such state denotes frequency (probability). In the context of symmetric games, when the equilibrium state is chosen as the ESS, the KL-divergence between the ESS and its neighbouring state diminishes over time in continuous-time replicator dynamics~\cite{baez_entropy}.
	
Another significance of the KL-divergence is that it serves as a Lyapunov function for the replicator dynamics. However, a crucial nuance comes into play: In two-strategy games, local asymptotic stability and evolutionary stability of a fixed point states are interchangeable, but this equivalence does not always hold in games featuring a larger number of strategies. In such instances, an ESS ensures the local asymptotic stability of the corresponding fixed point, while the converse assertion may not be valid~\cite{taylorjonker1978}.
	
On a much wider picture, the principle of relative entropy minimization, which is widely utilized in the realm of evolutionary dynamics~\cite{bomze1991cross, karev2010, baez_entropy}, states that when an evolving population state is in proximity to an equilibrium point, the relative entropy monotonically decreases. The application of this principle can be seen as an implementation of Kullback's proposal for the principle of minimum discrimination information, which has proven successful in various contexts \cite{karev2010, Floerchinger2020}, particularly in evolutionary systems.

Motivated by these findings, we now aim to extend this understanding to asymmetric games. {\color{black}An important point to note is that in this section (unlike the last section), we need not confine ourselves to the cases where each population only has two types:  \emph{Thus, in this section, henceforth, $\alpha>1$ and $\beta>1$, unless otherwise stated.}}
\subsection{Fixed Points}\label{sec:info_fixed}
Before examining the outcomes for a periodic orbit, let's first explore the fixed point, $(\mathbf{\hat{x}},\mathbf{\hat{y}})$, corresponds to the decrease of KL-divergence. As previously motivated, our primary attention is directed towards the replicator maps, which are represented by Eq.~(\ref{eqn:asymmetric_replicator_gen}).
We define a combined relative entropy as:
\begin{equation}\label{eqn:kl_divergence1}
	V(\mathbf{x},\mathbf{y})\equiv D_{KL}(\mathbf{\hat{x}}||{\bf x}^{k})+rD_{KL}(\mathbf{\hat{ y}}||{\bf y}^{k}),
\end{equation}
where $r$ is some positive constant, which acts to incorporate a weighting factor determining the degree of relevance attributed to the information provided by variable ${\bf x}$ vis-a-vis that by ${\bf y}$. It is very important to keep in mind the fact that KL-divergence and, hence, $V$ above is a non-negative quantity. Expanding R.H.S. of Eq.~(\ref{eqn:kl_divergence1}), we get
\begin{equation}\label{eqn:kl_divergence2}
	V(\mathbf{x},\mathbf{y})=-\sum_i\hat{x}_i\log\left(\frac{x_i}{\hat{x}_i}\right)-r\sum_i\hat{y}_i\log\left(\frac{y_i}{\hat{y}_i}\right).
\end{equation}
Now, we recall Jensen's inequality, which states that for a convex function $\phi$,~$\phi(\mathbb{E}[x])\leq \mathbb{E}[\phi(x)]$, where $\mathbb{E}[\phi]$ represents the expectation value of $\phi$. Given the fact that $-\log(\lambda)$ is a convex function of $\lambda$, we can now apply Jensen's inequality to Eq.~(\ref{eqn:kl_divergence2}) as follows:
\begin{eqnarray}
&&-\log(\sum_ix_i)-r\log(\sum_iy_i)\leq-\sum_i\hat{x}_i\log\left(\frac{x_i}{\hat{x}_i}\right)-r\sum_i\hat{y}_i\log\left(\frac{y_i}{\hat{y}_i}\right),\\
\implies&&0\leq-\sum_i\hat{x}_i\log\left(\frac{x_i}{\hat{x}_i}\right)-r\sum_i\hat{y}_i\log\left(\frac{y_i}{\hat{y}_i}\right),
\end{eqnarray}
where, obviously, at $(\mathbf{x},\mathbf{y})=(\mathbf{\hat{x}},\mathbf{\hat{y}})$, R.H.S. vanishes; therefore, at $(\mathbf{\hat{x}},\mathbf{\hat{y}})$ the function $V$ achieves its minimum value, viz., zero. 

Next, we define
\begin{equation}
\Delta	V(\mathbf{x},\mathbf{y})\equiv \Delta D_{KL}(\mathbf{\hat{x}}||\mathbf{{ x}}^{k})+r\Delta D_{KL}(\mathbf{\hat{y}}||{\bf y}^{k}),
\end{equation}
where
\begin{equation}
	\Delta D_{KL}(\mathbf{\hat{x}}||{\bf x}^{k})\equiv	D_{KL}(\mathbf{\hat{x}}||{\bf x}^{k+1})-D_{KL}(\mathbf{\hat{x}}||{\bf x}^k).
\end{equation}
Therefore, using the replicator equation (\ref{eqn:replicator2}), we arrive at
\begin{equation}
\Delta V=-\sum_{i=1}\hat{x}_i\log\left(1+\delta^k_{x_i}\right)-r\sum_{i=1}\hat{y}_i\log\left(1+\delta^k_{y_i}\right).
\end{equation}
Here $\delta^k_{x_i}\equiv ({\sf A}\mathbf{x}^k)_i+({\sf B} \mathbf{y}^k)_i-\mathbf{x}^k\cdot{\sf A}\mathbf{x}^k-\mathbf{x}^k\cdot{\sf B}\mathbf{y}^k$ and $\delta^k_{y_i}\equiv({\sf C}\mathbf{x}^k)_i+({\sf D}\mathbf{y}^k)_i-\mathbf{y}^k\cdot{\sf C}\mathbf{x}^k-\mathbf{y}^k\cdot{\sf D}\mathbf{y}^k$. Next, since `$-\log$' is the convex function, we use Jensen's inequality to get
\begin{equation}\label{eqn:delta_v_intermidiate}
\Delta V	\geq -\log\left[\sum_{i=1}\hat{x}_i(1+\delta^k_{x_i})\right]-r\log\left[\sum_{j=1}\hat{y}_i(1+\delta^k_{y_i})\right].
\end{equation}
Now we see that if $\Delta V<0$ then either
\begin{equation}
	0 > -\log\left[\left(1+\sum_{i=1}\hat{x}_i\delta^k_{x_i}\right)\right]~
\text{or}~ 0 > -\log\left[\left(1+\sum_{i=1}\hat{y}_i\delta^k_{y_i}\right)\right].
\end{equation}
This implies that
	\begin{eqnarray}\label{26}
		\mathbf{\hat{x}}\cdot{\sf A}\mathbf{x}^{k}+\mathbf{\hat{x}}\cdot{\sf B}\mathbf{y}^{k}>\mathbf{x}^k\cdot{\sf A}\mathbf{x}^{k}+\mathbf{x}^k\cdot{\sf B}\mathbf{y}^{k}
		~\text{or}~
		\mathbf{\hat{y}}\cdot{\sf C}\mathbf{x}^{k}+\mathbf{\hat{y}}\cdot{\sf D}\mathbf{y}^{k}>\mathbf{y}^k\cdot{\sf C}\mathbf{x}^{k}+\mathbf{y}^k\cdot{\sf D}\mathbf{y}^{k}.
	\end{eqnarray}
which is essentially the ESS condition. Thus, we conclude that decrease in combined relative entropy during the time-evolution leads to achieving ESS.
\subsection{Periodic Orbit}\label{sec:info_periodic}
In view of the connection between relative entropy and fixed point that is, ESS, an extension of the definition of ESS for the case of periodic orbits may be envisaged, we write the following for a sequence of states $\{(\mathbf{\hat{x}}^1,\mathbf{\hat{y}}^1),\cdots,(\mathbf{\hat{x}}^m,\mathbf{\hat{y}}^m)\}$:

\textbf{Definition 5a}: A sequence of states $\{(\mathbf{\hat{x}}^k,\mathbf{\hat{y}}^k):  k=1,2,\cdots,m\,{\rm and}\,(\mathbf{\hat{x}}^i,\mathbf{\hat{y}}^i) \ne (\mathbf{\hat{x}}^j,\mathbf{\hat{y}}^j)$, $\forall i\ne j\}$ is a  2ISO($m$) (two-species Information Stable Orbit)
of the replicator map for asymmetric games if either
\begin{subequations}\label{eqn:aiso1}
	\begin{eqnarray}
		&&\sum_{k=1}^{m}\mathbf{\hat{x}}^1\cdot{\sf A}\mathbf{x}^{k}+\mathbf{\hat{x}}^1\cdot{\sf B}\mathbf{y}^{k}>	\sum_{k=1}^{m}\mathbf{x}^{k}\cdot{\sf A}\mathbf{x}^{k}+\mathbf{x}^{k}\cdot{\sf B}\mathbf{y}^{k},\\
		\text{or} \nonumber &&\\
		&&\sum_{k=1}^{m}	\mathbf{\hat{y}}^1\cdot{\sf C}\mathbf{x}^{k}+\mathbf{\hat{y}}^1\cdot{\sf D}\mathbf{y}^{k}>	\sum_{k=1}^{m}\mathbf{y}^{k}\cdot{\sf C}\mathbf{x}^{k}+\mathbf{y}^{k}\cdot{\sf D}\mathbf{y}^{k};
	\end{eqnarray}
\end{subequations}
where $\{(\mathbf{x}^k,\mathbf{y}^k)\}_{k=1}^{k=m}$ is any sequence of $m$ distinct states of the map
starting in some infinitesimal deleted neighbourhood of $(\mathbf{\hat{x}}^1,\mathbf{\hat{y}}^1)$.

The 2ISO($m$) definition characterizes any finite sequence of $m$ states, denoted by $\{(\mathbf{\hat{x}}^{k},\mathbf{\hat{y}}^{k})\}_{k=1}^{k=m}$, of a map by comparing it with a corresponding sequence of $m$ mutant states, denoted by $\{(\mathbf{x}^{k},\mathbf{y}^{k})\}_{k=1}^{k=m}$, using Eq.~(\ref{eqn:aiso1}).
It's important to note that only the initial state $(\mathbf{\hat{x}}^{1},\mathbf{\hat{y}}^{1})$ of the given sequence is directly involved in the comparison, while subsequent elements are determined by the map itself.
For the sequence of mutant states, it starts from a neighborhood surrounding the initial element of the 2ISO sequence, and subsequent elements are also determined by the map's dynamics.

Note that, Eq.~(\ref{eqn:aiso1}) is valid for any sequence, $\{(\mathbf{\hat{x}}^{k},\mathbf{\hat{y}}^{k})\}_{k=1}^{k=m}$ and its neighborhood sequence $\{(\mathbf{x}^{k},\mathbf{y}^{k})\}_{k=1}^{k=m}$, hence it is blind to the notion of a periodic orbit. Therefore, to verify a periodic orbit's qualification as 2ISO($m$), it's imperative that each of the $m$ sequences, each of length $m$ starting from the $m$ distinct states of an $m$-periodic orbit, satisfies the 2ISO($m$) condition. Here, all the $m$ sequences denoted by $\{(\mathbf{\hat{x}}^{j},\mathbf{\hat{y}}^{j})\}_{j=k}^{j=k+m-1}$ represents the same $m$-period orbit. The corresponding sequences of neighboring points, denoted by  $\{(\mathbf{x}^{j},\mathbf{y}^{j})\}_{j=k}^{j=k+m-1}$ originate from the infinitesimal deleted neighbourhood of $(\mathbf{\hat{x}}^{k},\mathbf{\hat{y}}^{k})$. Therefore, we can rewrite the 2ISO condition (Eq.~(\ref{eqn:aiso1})) for an $m$ period orbit as follows:

\textbf{Definition 5b}: A sequence of states $\{(\mathbf{\hat{x}}^k,\mathbf{\hat{y}}^k): k=1,2,\cdots,m\,{\rm and}\,(\mathbf{\hat{x}}^i,\mathbf{\hat{y}}^i) \ne (\hat{\mathbf{x}}^j,\hat{\mathbf{y}}^j)$, $\forall i\ne j\}$ with period $m$, is a  2ISO($m$)
of the replicator map for asymmetric games if either
\begin{subequations}\label{eq:aiso2}
\begin{eqnarray}
	&&\sum_{j=k}^{k+m-1}\mathbf{\hat{x}}^k\cdot{\sf A}\mathbf{x}^{j}+\mathbf{\hat{x}}^k\cdot{\sf B}\mathbf{y}^{j}>	\sum_{j=k}^{k+m-1}\mathbf{x}^j\cdot{\sf A}\mathbf{x}^{j}+\mathbf{x}^j\cdot{\sf B}\mathbf{y}^{j},\\
\text{or} \nonumber &&\\
	&&\sum_{j=k}^{k+m-1}	\mathbf{\hat{y}}^k\cdot{\sf C}\mathbf{x}^{j}+\mathbf{\hat{y}}^k\cdot{\sf D}\mathbf{y}^{j}>	\sum_{j=k}^{k+m-1}\mathbf{y}^j\cdot{\sf C}\mathbf{x}^{j}+\mathbf{y}^j\cdot{\sf D}\mathbf{y}^{j};
\end{eqnarray}
\end{subequations}
where $\{(\mathbf{x}^j,\mathbf{y}^j)\}_{j=k}^{j=k+m-1}$ is any sequence of $m$ distinct states of the map starting in some infinitesimal deleted neighbourhood of $(\mathbf{\hat{x}}^k,\mathbf{\hat{y}}^k)$.

This definition makes sense if it can relate 2ISO for periodic orbits to the decrease in corresponding KL-divergence. To this end, we recall that $m$-period orbit points of a map are fixed points of $m$-th iterate of the map and introduce the notations---$\Delta_mD_{KL}(\mathbf{\hat{x}}^k||\mathbf{x}^k)\equiv D_{KL}(\mathbf{\hat{x}}^k||\mathbf{x}^{k+m})-D_{KL}(\mathbf{\hat{x}}^k|\mathbf{x}^k)$ and $\Delta_mD_{KL}(\mathbf{\hat{y}}^k||\mathbf{y}^k)\equiv D_{KL}(\mathbf{\hat{y}}^k||\mathbf{y}^{k+m})-D_{KL}(\mathbf{\hat{y}}^k||\mathbf{y}^k)$. We now start with,
\begin{subequations}
\begin{eqnarray}
&&\Delta_m V\equiv \Delta_mD_{KL}(\mathbf{\hat{x}}^k||\mathbf{x}^k)+r\Delta_mD_{KL}(\mathbf{\hat{y}}^k||\mathbf{y}^k).\\
\implies&&\Delta_m V=-\sum_i\hat{x}_i^k\log\left(\frac{x_i^{k+m}}{x_i^k}\right)-r\sum_i\hat{y}_i^k\log\left(\frac{y_i^{k+m}}{y_i^{k}}\right),\\
\implies&&\Delta_m V=-m\sum_i\hat{x}_i^k\sum_{j=k}^{k+m-1}\frac{1}{m}\log\left(1+\delta^j_{x_i}\right)-rm\sum_i\hat{y}_i^k\sum_{j=k}^{k+m-1}\frac{1}{m}\log\left(1+\delta^j_{y_i}\right),	\\
\implies&&\Delta_m V	\geq-m\sum_i\hat{x}_i^k\log\left(1+\frac{1}{m}\sum_{j=k}^{k+m-1}\delta^j_{x_i}\right)-rm\sum_i\hat{y}_i^k\log\left(1+\frac{1}{m}\sum_{j=k}^{k+m-1}\delta^j_{y_i}\right),\\
\implies&&\Delta_m V	\geq-m\log\left[\left(1+\frac{1}{m}\sum_i\hat{x}_i^k\sum_{j=k}^{k+m-1}\delta^j_{x_i}\right)\right]-rm\log\left[\left(1+\frac{1}{m}\sum_i\hat{y}_i^k\sum_{j=k}^{k+m-1}\delta^j_{y_i}\right)\right].
\end{eqnarray}
\end{subequations}
where the inequalities are due to Jensen's inequality. Thus, if $\Delta_m V<0$, then either
\begin{equation}
	0>-\log\left[\left(1+\frac{1}{m}\sum_i\hat{x}_i^k\sum_{j=k}^{k+m-1}\delta^j_{x_i}\right)\right]~
\text{or}~	0>-\log\left[\left(1+\frac{1}{m}\sum_i\hat{y}_i^k\sum_{j=k}^{k+m-1}\delta^j_{y_i}\right)\right].
\end{equation}
These conditions can be rewritten as either
\begin{equation}
	\sum_{j=k}^{k+m-1}\mathbf{\hat{x}}^k\cdot{\sf A}\mathbf{x}^{j}+\mathbf{\hat{x}}^k\cdot{\sf B}\mathbf{y}^{j}>	\sum_{j=k}^{k+m-1}\mathbf{x}^j\cdot{\sf A}\mathbf{x}^{j}+\mathbf{x}^j\cdot{\sf B}\mathbf{y}^{j},
\end{equation}
or,
\begin{equation}
	\sum_{j=k}^{k+m-1}	\mathbf{\hat{y}}^k\cdot{\sf C}\mathbf{x}^{j}+\mathbf{\hat{y}}^k\cdot{\sf D}\mathbf{y}^{j}>	\sum_{j=k}^{k+m-1}\mathbf{y}^j\cdot{\sf C}\mathbf{x}^{j}+\mathbf{y}^j\cdot{\sf D}\mathbf{y}^{j}.
\end{equation}
This is nothing but the definition (Eq.~(\ref{eq:aiso2})) of 2ISO. Thus, we conclude that decrease in $V$ during the time-evolution of replicator dynamics leads to achieving 2ISO.
\subsection{2ISO and dynamic stability}\label{iso_dynamics}
The other extension of ESS, viz., 2HSO($m$), corresponded to stable periodic orbits. Now, we wonder whether a similar relationship exists between periodic orbits and 2ISO. It appears to be so. However, providing a mathematical proof for periodic orbits of general periods is rather challenging. Nevertheless, we can rigorously establish the following proof for the two-period orbit in the case of {\color{black} two-strategy games ($\alpha=\beta=2$)}:

\textbf{Proposition 3:} A locally asymptotically stable 2-period orbit of a replicator map for two-player, two-strategy game is a 2ISO.

\emph{Proof:} The line of argument we take is that if we can prove that 2HSO($2$) implies 2ISO($2$), then we have effectively proven that local asymptotic stability implies 2ISO($2$) since we know that local asymptotic stability implies 2HSO($2$). We begin with expanding inequality~\ref{eqn:HSO}(a) with $m=2$ and recasting it as,
	\begin{equation}
\mathbf{\hat{x}}^1\cdot{\sf A}\mathbf{x}^1+\mathbf{\hat{x}}^1\cdot {\sf B}\mathbf{y}^1+\frac{H_{\mathbf{{x}}^2}}{H_{\mathbf{{x}}^1}}[\mathbf{\hat{x}}^1\cdot{\sf A}\mathbf{x}^2+\mathbf{\hat{x}}^1\cdot {\sf B}\mathbf{y}^2]>\mathbf{x}^1\cdot{\sf A}\mathbf{x}^1+\mathbf{x}^1\cdot {\sf B}\mathbf{y}^1+\frac{H_{\mathbf{{x}}^2}}{H_{\mathbf{{x}}^1}}[\mathbf{x}^1\cdot{\sf A}\mathbf{x}^2+\mathbf{x}^1\cdot {\sf B}\mathbf{y}^2].
	\end{equation}
We add $\mathbf{\hat{x}}^1\cdot{\sf A}\mathbf{x}^2+\mathbf{\hat{x}}^1\cdot {\sf B}\mathbf{y}^2-\mathbf{x}^2\cdot{\sf A}\mathbf{x}^2-\mathbf{x}^2\cdot {\sf B}\mathbf{y}^2$ on both the sides to get,
\begin{equation}\label{eq:in}
	\sum_{j=1}^{2}[\mathbf{\hat{x}}^1\cdot{\sf A}\mathbf{x}^{j}+\mathbf{\hat{x}}^1\cdot{\sf B}\mathbf{y}^{j}-\mathbf{x}^j\cdot{\sf A}\mathbf{x}^{j}-\mathbf{x}^j\cdot{\sf B}\mathbf{y}^{j}]>\frac{H_{\mathbf{{x}}^2}}{H_{\mathbf{{x}}^1}}[\mathbf{x}^1\cdot{\sf A}\mathbf{x}^2+\mathbf{x}^1\cdot {\sf B}\mathbf{y}^2-\mathbf{\hat{x}}^1\cdot{\sf A}\mathbf{x}^2-\mathbf{\hat{x}}^1\cdot {\sf B}\mathbf{y}^2]+\mathbf{\hat{x}}^1\cdot{\sf A}\mathbf{x}^2+\mathbf{\hat{x}}^1\cdot {\sf B}\mathbf{y}^2-\mathbf{x}^2\cdot{\sf A}\mathbf{x}^2-\mathbf{x}^2\cdot {\sf B}\mathbf{y}^2.
\end{equation}
Now to show that the 2ISO condition holds, one has to show that inequality (\ref{eq:in}) has non-negative R.H.S. which can be rearranged as,
\begin{equation}
	[({\sf A}\mathbf{x}^2)_1+({\sf B}\mathbf{y}^2)_1-({\sf A}\mathbf{x}^2)_2-({\sf B}\mathbf{y}^2)_2]\left[(\hat{x}^1-x^2)-\frac{H_{\mathbf{{x}}^2}}{H_{\mathbf{{x}}^1}}(\hat{x}^1-x^1)\right]
	=2(H_{\mathbf{{x}}^2})^{-1}(\Delta x^2)\left[(\hat{x}^1-x^2)-\frac{H_{\mathbf{{x}}^2}}{H_{\mathbf{{x}}^1}}(\hat{x}^1-x^1)\right]\label{HSO_implies_ISO},
\end{equation}
 where $\Delta x^2=x^3-x^2$. We are free to choose small enough neighbourhood of $\hat{x}^1$ so that $x^1$ is close enough such that the term containing $\hat{x}^1-x^1$ in the R.H.S. of Eq.~(\ref{HSO_implies_ISO}) can be made arbitrarily small (note that its prefactor composed of heterogeneity factors is finite). Furthermore, for stable periodic orbits and the aforementioned small neighbourhood, $ \Delta x^2$ and $\hat{x}^1-x^2$ should be of the same sign. In conclusion, the R.H.S. of Eq.~(\ref{HSO_implies_ISO}) is a positive quantity for some small neighbourhood of $\hat{x}^1$. Similar conclusion can be drawn, had we started with inequality~\ref{eqn:HSO}(b). Thus, we have proven that 2HSO($2$) implies 2ISO($2$), and hence, Proposition 3 is true.

{\color{black}What about unstable periodic orbits? They need not be ISO. As a trivial example, if
\begin{equation}
	{\sf A} = \begin{pmatrix}
		5 & 1 \\
		3 & 4 
	\end{pmatrix},
	~{\sf B} = \begin{pmatrix}
		6 & 1 \\
		3 & 4 
	\end{pmatrix},
	~{\sf C}=\begin{pmatrix}
		6 & 1 \\
		3 & 4 
	\end{pmatrix}~\text{and}~{\sf D}=\begin{pmatrix}
		5 & 1 \\
		3 & 4 
	\end{pmatrix}.
\end{equation}
then there is an unstable interior fixed point of the replicator map and it is not a 2ISO(1), i.e., a 2ESS. We are unable to comment something general about periodic orbits with arbitrary period. For 2-period orbit, the result is interesting, as encapsulated by the following proposition.
}

\textbf{Proposition 4:} A two-period orbit of the asymmetric replicator map for two-player, two-strategy game is a 2ISO.

\emph{Proof:} Let us consider a two-period periodic orbit, denoted as, $\{(\mathbf{\hat{x}}^{1},\mathbf{\hat{y}}^{1}),(\mathbf{\hat{x}}^{2},\mathbf{\hat{y}}^{2})\}$, we can easily see that $\Delta\hat{x}^{1}\Delta\hat{x}^2\leq0$ and $\Delta\hat{y}^{1}\Delta\hat{y}^2\leq0$. It is reasonable to assume that the two periodic points are sufficiently distant from each other and to choose the state $(\mathbf{x}^1,\mathbf{y}^1)$ within an infinitesimally small neighbourhood of $(\mathbf{\hat{x}}^1,\mathbf{\hat{y}}^1)$. Consequently, the inequality, $\Delta x^1\Delta x^2+r(\Delta y^1\Delta y^2)<0$, where $r$ is some positive number, holds true. This expectation remains valid even in the case of an unstable periodic orbit for sufficient closeness between  $(\mathbf{x}^1,\mathbf{y}^1)$ and  $(\mathbf{\hat{x}}^1,\mathbf{\hat{y}}^1)$. Consequently, we have
\begin{equation}\label{eqn:note1}
\Delta{x}^{1}\Delta{x}^2+\sum_{j=1}^2\left(\frac{H_{\mathbf{x}^2}}{H_{\mathbf{x}^j}}\right)\Delta x^j(x^1-\hat{x}^1)
+r\left(\Delta{y}^{1}\Delta{y}^2+\sum_{j=1}^2\left(\frac{H_{\mathbf{y}^2}}{H_{\mathbf{y}^j}}\right)\Delta y^j(y^1-\hat{y}^1)\right)<0.
\end{equation}
This is because, given the heterogeneity factors are positive quantities, the second and the fourth terms can be made arbitrarily small by taking $x^1\to\hat{x}^1$ and $y^1\to\hat{y}^1$. We can rewrite Eq.~(\ref{eqn:note1}) as:
\begin{equation}
H_{\mathbf{x}^2}\left[\sum_{j=1}^2(H_{\mathbf{x}^j})^{-1}\Delta x^j(x^j-\hat{x}^1)\right]+rH_{\mathbf{y}^2}\left[\sum_{j=1}^2(H_{\mathbf{y}^j})^{-1}\Delta y^j(y^j-\hat{y}^1)\right]<0.
\end{equation}
Using the asymmetric replicator map, we can rewrite it as,
\begin{equation}
	H_{\mathbf{x}^2}\sum_{j=1}^2(x^j-\hat{x}^1)[({\sf A}\mathbf{x}^j)_1+({\sf B}\mathbf{y}^j)_1-({\sf A}\mathbf{x}^j)_2-({\sf B}\mathbf{y}^j)_2]+rH_{\mathbf{y}^2}\sum_{j=1}^2(y^j-\hat{y}^1)[({\sf C}\mathbf{x}^j)_1+({\sf D}\mathbf{y}^j)_1-({\sf C}\mathbf{x}^j)_2-({\sf D}\mathbf{y}^j)_2]<0.
\end{equation}
This is equivalent to the 2ISO($2$) for 2-period orbits, as it implies that
\begin{equation}
	\sum_{j=1}^2\mathbf{x}^j\cdot{\sf A}\mathbf{x}^{j}+\mathbf{x}^j\cdot{\sf B}\mathbf{y}^{j}-\mathbf{\hat{x}}^1\cdot{\sf A}\mathbf{x}^{j}-\mathbf{\hat{x}}^1\cdot{\sf B}\mathbf{y}^{j}<0~\text{or}~\sum_{j=1}^2\mathbf{y}^j\cdot{\sf C}\mathbf{x}^{j}+\mathbf{y}^j\cdot{\sf D}\mathbf{y}^{j}-\mathbf{\hat{y}}^1\cdot{\sf C}\mathbf{x}^{j}-\mathbf{\hat{y}}^1\cdot{\sf D}\mathbf{y}^{j}<0.
\end{equation}
Therefore, we conclude that a 2-period orbit qualifies as 2ISO regardless of its nature of stability. 
\section{Numerical verification}\label{sec:numver}
Lastly, we do some numerical analysis to illustrate that the stable periodic orbit in fact corresponds to 2ISO and 2HSO. However, prior to delving into this, it is imperative to ensure that parameters are within a suitable range---strict physical region---in order to render the forward invariance of the asymmetric replicator map; i.e., starting from initial conditions within the range $(0,1)\times(0,1)$, the dynamics remain within this range at subsequent times, thereby ensuring that frequencies attain only physically meaningful values. While the required conditions on the parameter have been analytically found in Sec.~\ref{sec:tdre}, here we verify them numerically. 

\begin{figure}[h!]
	\includegraphics[scale=0.95]{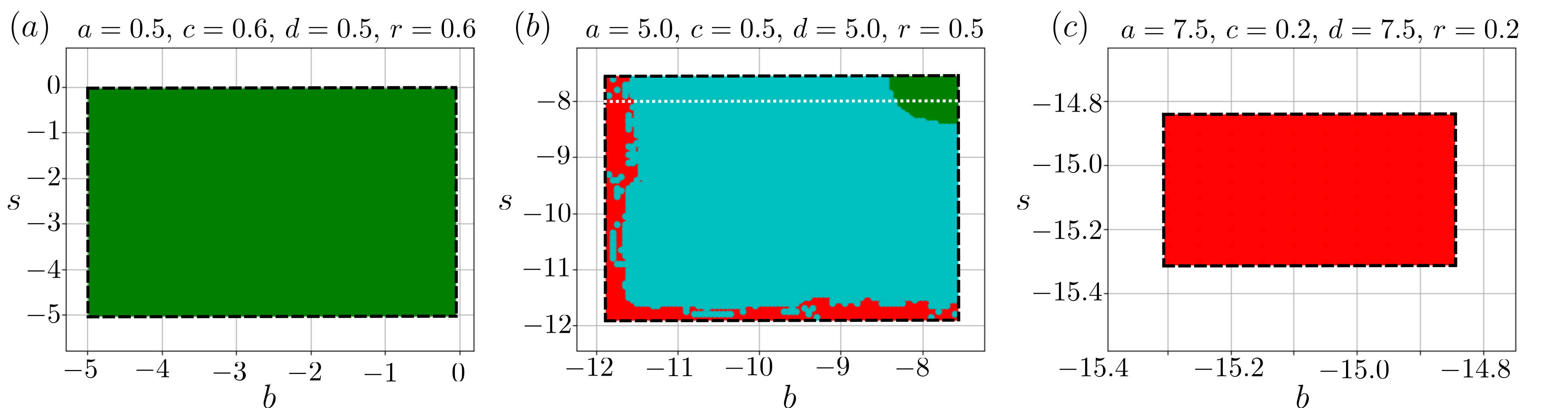}
	\caption{Strict physical region for asymmetric replicator map:  In the three subplots---viz., (a), (b) and (c)---parameters $b$ and $s$ are varied but other four parameters are kept fixed at what is written on top of respective figures. White colour indicates that for (at least some) 625 initial conditions at any of the point there, the $(x,y)$ values of the trajectories  leave the phase space. In contrast, non-white colour indicates that for all the 625 initial conditions at any of the point there, the $(x,y)$ values of the trajectories never leave the phase space. Black dashed lines demarcate the analytically found strict physical region (Sec.~\ref{sec:tdre}); the fact that these lines separate white from non-white regions validate the analytical arguments. Among the colours, green denotes (stable) fixed points, cyan denotes (stable) periodic orbits, and red denotes chaotic orbits. The white dotted line in subplot (b) is range of $b$ parameter over which we illustrate the bifurcation diagram in Fig.~\ref{fig:bifurcation}.}
	\label{fig:strict_physical_region}
\end{figure}

There are six parameters in all and so to illustrate them in two dimensions in this paper, we fix four of them (such that they are in analytically determined strict physical region) and vary the other two. For each fixed set of values of the parameters, the system is time-evolved for {625} initial conditions of $(x,y)$ picked uniformly from the square grid--$(0,1)\times(0,1)$ and it is checked whether all the trajectories reaches an attractor within physically allowed phase space or leaves the phase space. In order to resolve the possible confusion in differentiating between a periodic orbit with large period and chaotic orbit, we additionally calculate the maximum Lyapunov exponent~\cite{Argyris_book,wiggins_2003}, which should be positive only for the latter. The strict physical region is illustrated in Fig.~\ref{fig:strict_physical_region}.

Now we can confidently pick parameter values such that we can get stable periodic orbits to check for their being 2HSO or 2ISO. For this purpose, it helps to note that \emph{neighbouring trajectory} $\{({\bf x}^{k},{\bf y}^{k})\}_{k=1}^{k=m}$ of  an $m$-period orbit as $\{(\mathbf{\hat{x}}^{k},\mathbf{\hat{y}}^{k})\}_{k=1}^{k=m}$ essentially starts at $(x^{1},y^{1})=(\hat{x}^{1}+\epsilon_x,\hat{y}^{1}+\epsilon_y$) where $\epsilon_x$ and $\epsilon_y$ are small real numbers. So, the 2ISO condition, Eq.~(\ref{eq:aiso2}),  may be compactly written as $\mu_x^m(\mathbf{\hat{x}}^{k},\epsilon_x,\epsilon_y)>0$ or $\mu_y^m(\mathbf{\hat{y}}^{k},\epsilon_x,\epsilon_y)>0$, where
\begin{subequations}\label{eqn:ISO_num}
\begin{eqnarray}
	\mu_x^m(\mathbf{\hat{x}}^{k},\epsilon_x,\epsilon_y)\equiv\sum_{j=k}^{k+m-1}\mathbf{\hat{x}}^k\cdot{\sf A}\mathbf{x}^{j}+\mathbf{\hat{x}}^k\cdot{\sf B}\mathbf{y}^{j}-\mathbf{x}^j\cdot{\sf A}\mathbf{x}^{j}-\mathbf{x}^j\cdot{\sf B}\mathbf{y}^{j},\label{eq:fig1}\\
\mu_y^m(\mathbf{\hat{y}}^{k},\epsilon_x,\epsilon_y)\equiv\sum_{j=k}^{k+m-1}	\mathbf{\hat{y}}^k\cdot{\sf C}\mathbf{x}^{j}+\mathbf{\hat{y}}^k\cdot{\sf D}\mathbf{y}^{j}	-\mathbf{y}^j\cdot{\sf C}\mathbf{x}^{j}-\mathbf{y}^j\cdot{\sf D}\mathbf{y}^{j}.
\end{eqnarray}
\end{subequations}
Thus, numerically it suffices to show that the above condition is satisfied for a range of small $\epsilon$-values in order for respective periodic orbits to be 2ISO. \textcolor{black}{Below}, we show this for two and four-period orbits. Note that even though our proof in the previous section is valid for two-period orbits only, but numerically we observe it to be valid for higher-period orbits. Similarly, we can verify that a particular periodic orbit is 2HSO by examining the positivity of any one of the following quantities [recall Eq.~(\ref{eqn:HSO})]:
\begin{subequations}\label{eqn:HSO_num}
\begin{eqnarray}
	\nu_x^m(\mathbf{\hat{x}}^{k},\epsilon_x,\epsilon_y)\equiv	\sum_{j=k}^{k+m-1} H_{\mathbf{{x}}^j}\left[ \mathbf{\hat{x}}^k\cdot{\sf A} \mathbf{{x}}^j +\mathbf{\hat{x}}^k\cdot{\sf B} \mathbf{{y}}^j-\mathbf{{x}}^k\cdot{\sf A} \mathbf{{x}}^j-\mathbf{{x}}^k\cdot{\sf B} \mathbf{{y}}^j \right],\\
	\nu_y^m(\mathbf{\hat{y}}^{k},\epsilon_x,\epsilon_y)\equiv\sum_{j=k}^{k+m-1} H_{\mathbf{{y}}^j} \left[\mathbf{\hat{y}}^k\cdot{\sf C} \mathbf{{x}}^j+\hat{\mathbf{y}}^k\cdot{\sf D} \mathbf{y}^j -\mathbf{{y}}^k\cdot{\sf C} \mathbf{{x}}^j-\mathbf{{y}}^k\cdot{\sf D} \mathbf{{y}}^j \right],
\end{eqnarray}
\end{subequations}
\textcolor{black}{which is also checked below.}

{\color{black}
We start our numerical quest by providing examples of well-known games and applying our definitions to these specific cases. As we already know that the system comprises four payoff matrices, each with its own significance (e.g., see \cite{hummert2014molbiostat}): Matrices ${\sf A}$ and ${\sf D}$ represent intraspecific competition within populations 1 and 2, respectively; while matrices ${\sf B}$ and ${\sf C}$ describe the interactions between the two populations, that is, interspecific interaction. It is important to keep in mind that the payoff elements are selected such that the parameters $a,~b,~c,~d,~r$ and $s$ must remain within the strict physical region.
	\begin{enumerate}
		\item[]\textbf{Example 1:} Consider a case where the games corresponding to intraspecific interaction are leader games and games corresponding to interspecific interaction are prisoner's dilemma game and chicken game, as represented by the following payoff matrices:
		\begin{equation}
			{\sf A} = \begin{pmatrix}
				1 & 3.2 \\
				7 & 0 
			\end{pmatrix},
			~{\sf B} = \begin{pmatrix}
				1 & -0.1 \\
				1.01 & 0 
			\end{pmatrix},
			~{\sf C}=\begin{pmatrix}
				1 & 0.1 \\
				1.01 & 0 
			\end{pmatrix}~\text{and}~{\sf D}=\begin{pmatrix}
				1& 3.2 \\
				7 & 0 
			\end{pmatrix}.
		\end{equation}
We find that the system has a single mixed Nash equilibrium, which corresponds to the interior fixed point of the replicator map [Eq.~(\ref{eqn:asymmetric_replicator_gen})], $(\hat{x}, \hat{y}) \approx (0.34, 0.35)$. This internal fixed point is not a 2ESS neither it is dynamically stable. Does that mean that the system has no stable outcome at all? Will the system reach  somewhere asymptotically with time? Fortunately, the answer to the first question is no, the system has a stable two period orbit, $\{p_1,p_2\}\approx\{(0.40,0.26),(0.28,0.42)\}$, as shown in the time series of Fig.~\ref{fig:illustration1}(a). Now let's check whether this periodic orbit satisfy the definitions of  HO($2$), 2HSO($2$), and 2ISO($2$). Since it is a two-period orbit, it must satisfy HO($2$) criteria by definition, which can be trivially verified by using Eq.~(\ref{21}). To check the 2HSO and 2ISO criteria, we plot $\nu_x^2$ and $\mu_x^2$ (see Eqs.~(\ref{eqn:ISO_num}) and (\ref{eqn:HSO_num})) in the neighbourhood of the corresponding periodic orbits, Fig.~\ref{fig:illustration1}(a). We can clearly see from the colorbar that it satisfies both 2HSO and 2ISO definitions.
		
\item[]\textbf{Example 2:} Next consider a scenario where the games corresponding to intraspecific interaction are anticoordination games and games corresponding to interspecific interaction are harmony games as depicted by the following payoff matrices:
		\begin{equation}
			{\sf A} = \begin{pmatrix}
				1.9 & 7.5 \\
				8.5 & 2.7 
			\end{pmatrix},
			~{\sf B} = \begin{pmatrix}
				1.8 & 1.2 \\
				1.1 & 1 
			\end{pmatrix},
			~{\sf C}=\begin{pmatrix}
				2.8 & 3.0 \\
				0.3 & 1 
			\end{pmatrix}~\text{and}~{\sf D}=\begin{pmatrix}
				0.0& 7.5 \\
				5.0 & 4.5 
			\end{pmatrix}.
		\end{equation}
		This system also has one mixed Nash equilibrium which is the interior fixed point, $(\hat{x},\hat{y})\approx(0.46,0.65)$. The internal fixed point is not a stable fixed point nor it is a 2ESS. Once again, we find ourselves in a situation where, if we consider only the fixed point, we might mistakenly conclude that the system lacks game-theoretic equilibrium. However, this is not the case, actually this system exhibits a stable four-period orbit, $\{q_1,q_2,q_3,q_4\}\approx \{(0.576,0.575),(0.262,0.743),(0.723,0.587),(0.132,0.747)\}$, as shown in the time series in Fig.~\ref{fig:illustration1}(b). As before, HO($4$) can be trivially verified by using Eq.~(\ref{21}). To check the 2HSO and 2ISO criteria, we plot $\nu_y^4$ and $\mu_x^4$ (see equations~(\ref{eqn:ISO_num}) and (\ref{eqn:HSO_num})) in the neighbourhood of the corresponding periodic orbits, Fig.~\ref{fig:illustration1}(b). We can clearly see from the colorbar that it satisfies both 2HSO and 2ISO criteria.
		\begin{figure}[h!]
			\hspace*{-0.2cm}\includegraphics[scale=0.65]{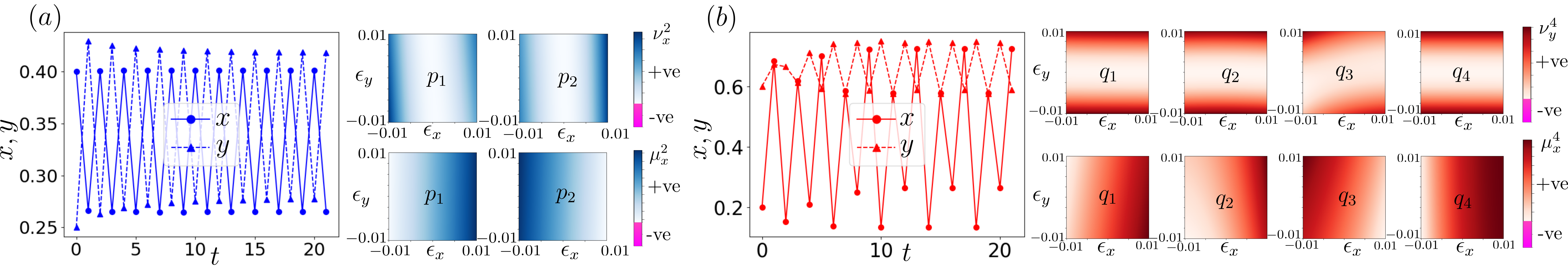}
			\caption{\textcolor{black}{Illustrative example showing that stable periodic orbits in discrete time asymmetric replicator dynamics is 2HSO and 2ISO: Subplot $(a)$ presents the time series and the fulfillment of the 2HSO and 2ISO conditions for Example 1. Similarly, subplot $(b)$ shows the time series and fulfillment of the 2HSO and 2ISO conditions for Example 2. Note that we do not explicitly display the values in the color bars; rather, we highlight the positivity and negativity of $\nu_x^2$, $\mu_x^2$, $\nu_y^4$ and $\mu_x^4$, as this is what matters in this context. The labels inside the colored boxes---i.e., $p_1,\,p_2,\,q_1,\,q_2,\,q_3$ and $q_4$---indicate the periodic orbit point, whose neighbourhood points serve as the starting points for the neighbouring trajectories. Time series for $x$ and $y$ are represented with solid (with circular marker) and dashed lines (with triangular marker), respectively. } }
			\label{fig:illustration1}
		\end{figure} 
		\item[]\textbf{Example 3:} Let us now consider the case where the games corresponding to intraspecific interaction are anticoordination games and the games corresponding to interspecific interaction are coordination games, as exemplified by the following payoff matrices:
		\begin{equation}
			{\sf A} = \begin{pmatrix}
				1 & 7 \\
				6 & 3 
			\end{pmatrix},
			~{\sf B} = \begin{pmatrix}
				4 & 2.5 \\
				2.5 & 3 
			\end{pmatrix},
			~{\sf C}=\begin{pmatrix}
				4 & 2.5 \\
				2.5 & 3 
			\end{pmatrix}~\text{and}~{\sf D}=\begin{pmatrix}
				1& 7 \\
				6 & 3 
			\end{pmatrix}.
		\end{equation}
		This system has one interior fixed point $(\hat{x},\hat{y})=(0.5,0.5)$ of the replicator map which is also only Nash equilibrium of the system. We also check that this internal fixed point is a 2ESS but it is not stable. However, unlike the other two examples, we find that it has two distinct stable two period orbits, $\{p_1,p_2\}\approx \{(0.80,0.31),(0.31,0.80)\}$ and  $\{p'_1,p'_2\}\approx\{(0.19,0.68),(0.68,0.19)\}$, as shown in the time series in Fig.~\ref{fig:illustration2}(a) and~\ref{fig:illustration2}(b) respectively, and depending on the initial conditions, the system reaches at either one of them. Again checking it is easy to see that they are HO($2$). To check the 2HSO and 2ISO criteria, we plot $\nu_x^2$ and $\mu_x^2$ in the neighbourhood of the corresponding periodic orbits, see Fig.~\ref{fig:illustration2}(a) and Fig.~\ref{fig:illustration2}(b). As we clearly see from the colorbar, both the periodic orbits are, in fact, 2HSO and 2ISO.
		\begin{figure}[h!]
			\includegraphics[scale=0.7]{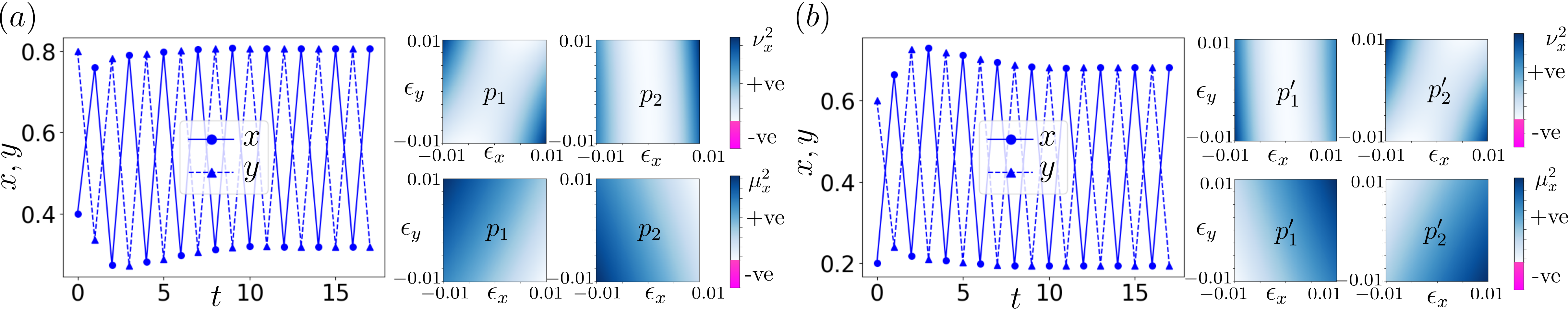}
			\caption{\textcolor{black}{Illustrative example demonstrating that stable periodic orbits in discrete time asymmetric replicator dynamics is 2HSO and 2ISO: Subplot $(a)$ and $(b)$ shows the time series and the validity of 2HSO and 2ISO conditions for the two different periodic orbits in Example 3. Once again, the labels $p_1,\,p_2,\,p'_1$ and $p'_2$ represent the periodic orbit points, with their neighbourhood points serving as the starting points for the neighbouring trajectories.}}
			\label{fig:illustration2}
		\end{figure}	
	\end{enumerate}
	
Through the above examples, we have highlight a crucial point that stable fixed points are not the only type of equilibria that can persist in a population. Periodic orbits, possessing game-theoretic interpretations, offer another viable possibility. While a single two-period orbit, a single four-period orbit, and two distinct coexisting two-period orbits have been illustrated by the examples, more rich dynamics is possible. To explore these further, we create a bifurcation diagram for a range of parameter values (see Fig.~\ref{fig:bifurcation}), revealing a variety of periodic orbits (and even chaotic orbit). However, we remind ourselves that while addressing evolutionary stability in such cases, one must consider extensions of the 2ESS concept, viz., 2HSO and 2ISO, since the traditional ESS framework does not apply to periodic orbits.

Before concluding this section, we would like to make a final remark: in Fig.~\ref{fig:bifurcation}, we observe the occurrence of a chaotic orbit as well (see Fig.~\ref{fig:bifurcation}(a)$(iv)$). Chaos has been noted in an evolutionary context before. Sato et al.~\cite{Sato2002} demonstrated the possibility of chaos in three-player learning games, while Rand et al.~\cite{Rand_1994} introduced the concept of the invasion exponent to study chaos in evolutionary models. The question remains: can we develop a game-theoretical understanding of chaos? While there have been some attempts in symmetric games~\cite{archan_chaos}, this remains an open avenue for further research in the context of asymmetric games.}

\begin{figure}[h!]
	\hspace*{-1cm}
	\centering
	\includegraphics[scale=0.75]{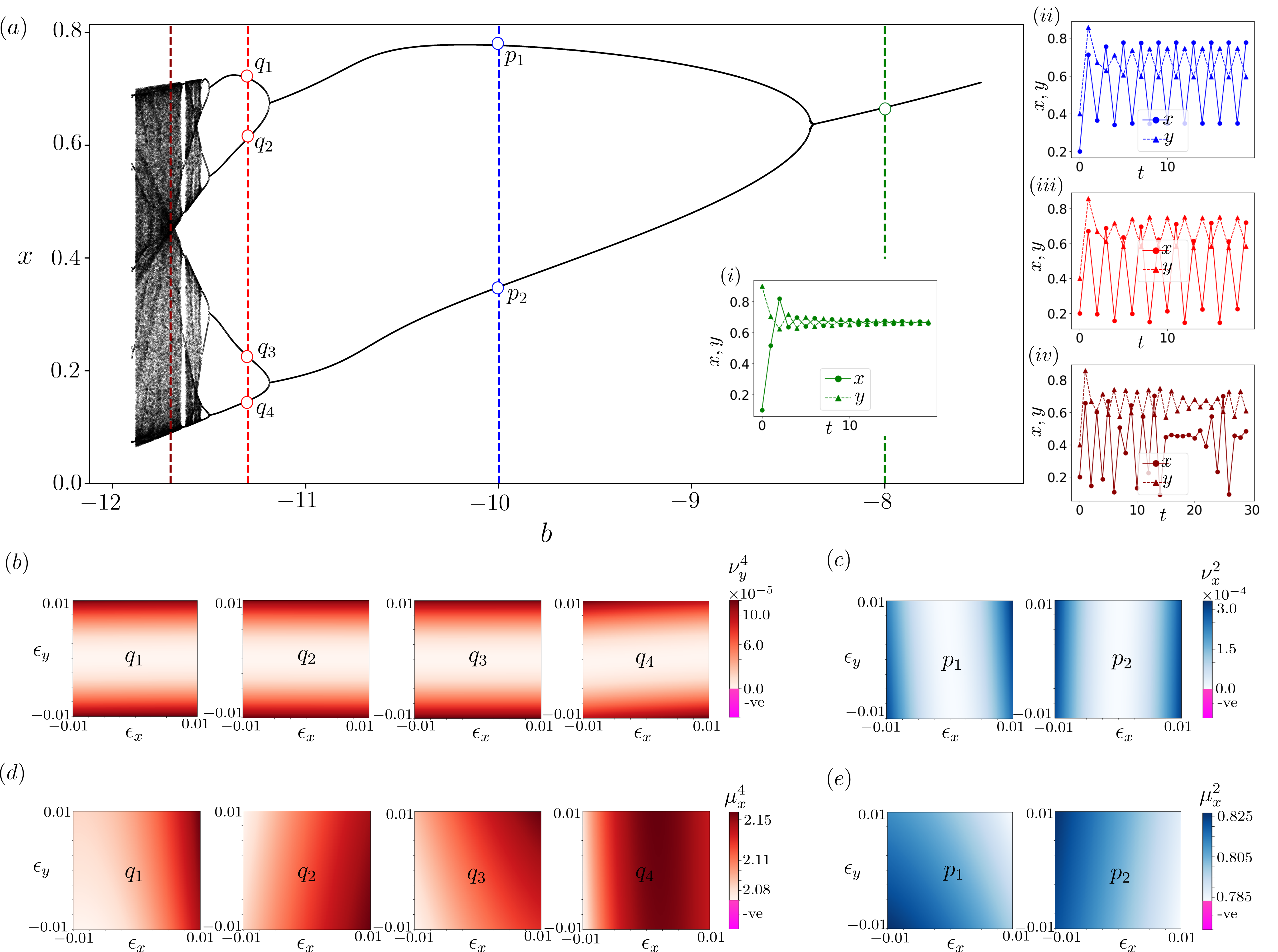}
	\caption{Stable periodic orbits of the discrete-time asymmetric replicator dynamics are 2HSO and 2ISO: Subplot (a) presents the bifurcation diagram of the asymmetric replicator map [Eq.~(\ref{eqn:replicator2by2})]. The varying parameter values forms white dotted line of Fig.~\ref{fig:strict_physical_region}(b), i.e., $a = 5.0$, $c = 0.5$, $d = 5.0$, $r = 0.5$, and $s = -8.0$, with $b$ ranging from $-12$ to $-7$. \textcolor{black}{At $b=-8.0$, the green dashed line intersects the bifurcation diagram at stable fixed point, $(\hat{x},\hat{y})\approx(0.66,0.66))$, as shown in the time series in subplot $(a)(i)$}. At $b = -10.0$. The blue dashed line intersects the bifurcation diagram at two stable 2-period orbit points, $\{p_1, p_2\}\approx\{(0.77,0.59), (0.34,0.74)\}$, \textcolor{black}{shown in the time series in subplot $(a)(ii)$}. Similarly, at $b = -11.3$, the red dashed line intersects the bifurcation diagram at four stable 4-period orbit points, $\{q_1, q_2, q_3, q_4\}\approx\{(0.71,0.58), (0.61,0.57), (0.22,0.74) (0.14,0.75)\}$, \textcolor{black}{whose time series is depicted in subplot $(a)(iii)$. Finally, at $b=-11.7$ the dark-red dashed line intersects the bifurcation diagram at a chaotic orbit, whose time series is shown in subplot $(a)(iv)$}. Subplots $(b)$ and $(c)$ illustrate the positivity of $\nu_y^4$ and $\nu_x^2$ for small values of $\epsilon$, respectively, affirming that the four-period and the two-period orbits are 2HSO. Likewise, subplots $(d)$ and $(e)$ depict the positivity of $\mu_x^4$ and $\mu_x^2$, respectively, validating that the four-period and the two-period orbits are 2ISO. \textcolor{black}{The labels $p_1$ and $p_2$ are periodic orbit points of two period orbit and $\,q_1,\,q_2,\,q_3$ and $q_4$ are periodic orbit points of the four period orbits whose neighbourhoods are the starting point of the corresponding neighbouring trajectories. }}
	\label{fig:bifurcation}
\end{figure}

\section{discussion and conclusion}\label{sec:VI}
The main goal of this paper has been on gaining insights about the oscillatory outcomes in asymmetric games through the combined lenses of evolutionary game theory and information theory. Specifically, we have put forward extensions of the usual concept of evolutionary stability for periodic orbits in discrete-time replicator dynamics. The extensions have been termed two-species heterogeneity stable orbit (2HSO) and two-species information stable orbit (2ISO). In the former, the fitness weighted by heterogeneity is optimized during the replication-selection process, while in the latter, fitness itself is optimized directly. Furthermore, while the former puts focus on the possible importance of  heterogeneity---the probability that two arbitrarily chosen members of the population belong to two
different phenotypes---in evolutionary dynamics in populations, the latter highlights a promising avenue of research into other information-theoretic ideas (e.g., R\'enyi entropy, Fisher information, and information geometry) in the context of evolutionary game theory. In either case, the local asymptotic stability of the periodic orbit get related to 2HSO and 2ISO. 

Furthermore, since HSO relates to the concept of strong stability~\cite{hofbauer_book,archan_periodic} in symmetric games, similar connection has been shown to be present between 2HSO and strongly stable strategy set in asymmetric games. This is a very much desirable connection given that strong stability validates the concept of evolutionary stability~\cite{hofbauer_book}. In the specific case of locally asymptotically stable $2$-period orbit, this connection carries over to 2ISO as well. 

It must be borne in mind that there are other game dynamics where oscillatory solutions are present, e.g., Brown--von Neumann--Nash dynamics and Smith dynamics possess limit cycles~\cite{Hofbauer2011}; Brown--von Neumann--Nash dynamics~\cite{Waters2009} even shows chaos.  The noisy version of best response dynamics~\cite{Ferraioli2013}, viz., logit dynamics, can lead to periodic outcomes.  In fact, replicator dynamics is a variant of incentive dynamics~\cite{Harper2014} whose other variants in discrete-time can show periodic orbits as well. The extension of NE or ESS (2ESS in the case of asymmetric games) for these other game dynamics, in principle, can be done as well by adopting case-by-case modification of the quantity to be optimized. 

Two interesting observations are worth pointing out. Firstly, our extensive numerical search for any periodic orbit in bimatrix games (truly asymmetric games) ended up in a failure. However, we could not provide a formal analytical proof for the absence of periodic orbits in bimatrix games. Secondly, we have already seen that a stable periodic orbit can be connected with the concept of 2HSO and 2ISO, but sometimes an unstable orbit can also be 2HSO or 2ISO. For 2ISO, we could provide proof for the connection only for 2-period orbits. In fact, the proof could be extended to show that all 2-period orbits---irrespective of their nature of stability---are 2ISO. However, this connection may not be independent of the nature of stability for other orbits with different periods---a conjecture which we could not prove. Going further, one may also pursue as a future problem, how chaotic oscillations in asymmetric game dynamics can be given both game-theoretic~\cite{archan_chaos} and information-theoretic meanings.

Nature is complex and so is its dynamics. We conclude by emphasizing that our work with replicator dynamics may be seen as an illustration of our bigger conviction that concentrating on game-theoretic meaning of only convergent outcomes in evolutionary dynamics may restrict us from exploring many hitherto undiscovered three-way connections between game theory, information theory, and dynamical systems theory. In future, we plan to extend the ideas developed herein to finite populations \cite{nowak2004emergence} and {\color{black} also to the case of continuous-time dynamics, probably through corresponding Poincar\'e maps. Besides exclusively restricting ourselves to the cases involving only two roles in this paper, we have presented some of the results only for the cases involving two strategies in both subpopulations. Extending such results to accommodate for an arbitrary number of roles and strategies is another potential direction for future work.}\\\\

\noindent
\textbf{Acknowledgements}
Sagar Chakraborty acknowledges the support from SERB (DST, govt. of India) through project no. MTR/2021/000119.\\\\
\textbf{Author Contributions} Vikash Kumar Dubey and Suman Chakraborty: Conceptualization, methodology, formal analysis, prepared figures, and writing-original draft.
Sagar Chakraborty: Conceptualization, methodology, validation, writing-original draft, and supervision.
All authors reviewed the manuscript.\\\\
\textbf{Funding}~~SERB (DST, govt. of India).\\\\
\textbf{Data availability statement}~~No data is associated with the manuscript.\\\\\\
{\Large{\textbf{Declarations}}}\\\\
\textbf{Conflict of interest}~~The authors have no conflicts of interest to declare.\\\\
\textbf{Ethical Approval}~~Not applicable. 
\bibliography{Dubey_etal_bibliography.bib}	
\end{document}